\documentclass[
reprint,
superscriptaddress,
amsmath,amssymb,
prx,
]{revtex4-2}
\usepackage[english]{babel}
\usepackage{graphicx}
\usepackage{dcolumn}
\usepackage{hyperref}
\usepackage{comment}
\usepackage{xcolor}
\usepackage{bm}
\newcommand*\chem[1]{\ensuremath{\mathrm{#1}}}

\begin{document}

\title{Fermi Surface and Lifshitz Transitions of a Ferromagnetic Superconductor under External Magnetic Fields}
	
\author{R. Leenen}
\affiliation{High Field Magnet Laboratory (HFML-EMFL), Radboud
University, Toernooiveld 7, 6525 ED Nijmegen, The Netherlands.}
\affiliation{Radboud University, Institute for Molecules and Materials, Heyendaalseweg 135, 6525 AJ Nijmegen, The Netherlands.}		
\author{D. Aoki}
\affiliation{Institute for Materials Research, Tohoku University, Ibaraki 311-1313, Japan}
\author{G. Knebel} 
\affiliation{Univ. Grenoble Alpes, CEA, Grenoble INP, IRIG, PHELIQS, F-38000 Grenoble, France}
\author{A. Pourret} 
\affiliation{Univ. Grenoble Alpes, CEA, Grenoble INP, IRIG, PHELIQS, F-38000 Grenoble, France}
\author{A. McCollam}
\email[Corresponding author:]{alix.mccollam@ru.nl}
\affiliation{High Field Magnet Laboratory (HFML-EMFL), Radboud
University, Toernooiveld 7, 6525 ED Nijmegen, The Netherlands.}
\affiliation{Radboud University, Institute for Molecules and Materials, Heyendaalseweg 135, 6525 AJ Nijmegen, The Netherlands.}

\begin{abstract}

Lifshitz transitions are being increasingly recognised as significant in a wide variety of strongly correlated and topological materials, and understanding the origin and influence of Lifshitz transitions is leading to deeper understanding of key aspects of magnetic, transport or quantum critical behaviour. 
In the ferromagnetic superconductor UCoGe,  a magnetic field applied along the $c$-axis has been shown to induce a series of anomalies in both transport and thermopower that may be caused by Lifshitz transitions.  The need to understand 
the subtleties of the relationship between magnetism, superconductivity and a heavy electron Fermi surface in the ferromagnetic superconductors makes it important to explore if and why a series of magnetic-field-induced Lifshitz transitions occurs in UCoGe.
Here we report magnetic susceptibility measurements of UCoGe, performed at temperatures down to 45 mK and magnetic fields ($\mu_0H ||c)$ up to 30 T.  We observe a series of clearly-defined features in the susceptibility, and multiple sets of strongly field-dependent de Haas-van Alphen oscillations, from which we extract detailed field-dependence of the quasiparticle properties. We complement our experimental results with density functional theory bandstructure calculations, and include a simple model of the influence of magnetic field on the calculated Fermi surface. By comparing experimental and calculated results, we determine the likely shape of the Fermi surface and identify candidate Lifshitz transitions that could correspond to two of the features in susceptibility. We connect these results to the development of magnetization in the system.
\end{abstract}

\maketitle

\section{Introduction}
Uranium-based intermetallics have partially filled 5\textit{f} electron orbitals that give rise to competition between Kondo hybridization, leading to heavy fermions, and RKKY interaction, leading to magnetic ordering. A small subset of the uranium heavy fermion systems are the ferromagnetic superconductors. \chem{UGe_2} was the first material discovered to support coexistence of ferromagnetism and superconductivity, under applied pressure between $p$ = 1.1~GPa and 1.6~GPa\cite{Saxena}. Later, coexistence of ferromagnetism and superconductivity was found in URhGe\cite{Aoki2001} and UCoGe\cite{Huy} at ambient pressure. The recent discovery of remarkable high-field re-entrant superconductivity in \chem{UTe_2} in the presence of ferromagnetic flutuations has added further interest to this research field \cite{Ran, Aoki2019, Aoki2022_JPCM, Rosuel2023}. 

In this article, we focus on UCoGe, which is a weak, uniaxial ferromagnet with a moment of $\sim 0.06\mu_B$ along the easy $c$-axis, and a Curie temperature of $T_C=2.7$~K \cite{Aoki2019}. Superconductivity coexists with ferromagnetism below a transition temperature of $T_{sc}=0.5$~K\cite{Huy, Huy2008}. Previous work showed that superconductivity is strongly influenced by ferromagnetic fluctuations associated with the 5$f$ electrons \cite{Karube, Hattori2012}, leading, for example, to highly anisotropic superconducting properties \cite{Aoki2009}. 
Because the 5\textit{f} and conduction electrons are hybridized into the Fermi sea, knowledge of the Fermi surface is highly desirable as a source of information about the magnetic and superconducting quasiparticles in 
this and other ferromagnetic or nearly ferromagnetic superconductors \cite{Sherkunov2018, McCollam2021, Aoki2022}.

Quantum oscillation studies are a powerful way to probe the Fermi surface.
Previous magnetoresistance and thermopower measurements of UCoGe showed several quantum oscillation frequencies, and identified a series of magnetic-field-induced features, proposed to be Lifshitz transitions \cite{Aoki2011,Malone2012, Bastien2016}, that is, topological changes of the Fermi surface. Lifshitz transitions \cite{Lifshitz1960} are
known to be important in heavy fermion systems, where they can be associated with a variety of phenomena, such as van Hove singularities \cite{Sherkunov2018, Hackl,Pfau2017, Pourret2019, McCollam2021}, or localization of $f$ electrons \cite{Pfau2013, Paschen2004, Hartmann2009, Shishido2005,Jiao2019, Mishra2021}, 
and often drive definitive features of magnetic, superconducting and quantum critical behavior \cite{Shishido2005, Park2006, Si2001, Si2014, Holmes2004, Rourke2008, Daou2006, Paschen2021}.
As well as a general knowledge of the Fermi surface of UCoGe, it is therefore also crucial to understand the nature of any observed Lifshitz transitions and the magnetic field evolution of the Fermi surface.  

Here we report on measurements of a$.$c$.$ magnetic susceptibility ($\chi$) and the de Haas-van Alphen (dHvA) effect in UCoGe, in magnetic fields parallel to the crystal $c$-axis.
We see six clear features in susceptibility that correspond to the previously proposed Lifshitz transitions \cite{Aoki2011, Bastien2016}. We also observe multiple, well-resolved dHvA oscillation frequencies.

As a complement to our experimental results, we performed density functional theory (DFT) calculations to determine the bandstructure and expected Fermi surface of UCoGe. We included a simple model of the spin-splitting effect of a magnetic field in our calculations.
By comparing the measured and predicted dHvA frequencies, we  are able to determine the likely shape of the Fermi surface. By examining the magnetic field dependence of the Fermi surface, we identify possible Lifshitz transitions consistent with the experimentally observed features in magnetic susceptibility.  
We suggest that only two of the observed features are due to Lifshitz transitions.

\section{Methods}
UCoGe crystallizes in the orthorhombic crystal structure of the TiNiSi type (space group {\em Pnma}). The high-quality single crystals used for this work were grown in a tetra-arc furnace using the Czochralski method \cite{Aoki2011b}. 
We measured two samples from different growth batches. Sample 1 (S1) has a residual resistance ratio (RRR = $R(300K)/R(\xrightarrow{}0K)$) of 110, and dimensions of $0.3$~mm~$\times~0.5$~mm~$\times~2.0$~mm in the \textit{a} $\times$ \textit{b} $\times$ \textit{c} directions, respectively. Sample 2 (S2) has RRR $\sim 40$ and 
dimensions of $1.0$~mm~$\times~1.0$~mm~$\times~1.4$~mm along the \textit{a} $\times$ \textit{b} $\times$ \textit{c} directions.

We performed magnetic susceptibility experiments up to 30~T, at temperatures down to 45~mK in a dilution refrigerator at the High Field Magnet Laboratory (HFML) in Nijmegen. We used the field modulation method, with modulation fields between $1.95$ and $3.51$~mT at frequencies between 100 and 150~Hz, with the modulation field and the d.c. field parallel to the $c$-axis of the sample.  
For the DFT calculations we used the APW+lo program WIEN2k\cite{Blaha}. We performed calculations using the GGA-PBE potential\cite{Perdew} and included spin-orbit coupling.  All uranium 5\textit{f} electrons were modelled as itinerant, that is they contributed to the Fermi volume. We used 30000 $k$-points to ensure good resolution at the Fermi energy, and from the calculated Fermi surface we could extract the expected dHvA frequencies\cite{Rourke} (see Supplementary Information). The lattice parameters were taken from a room temperature x-ray refinement study of a single crystal \cite{Samsel-Czekala}. 

\section{Results and discussion}
\subsection{Experimental results}

\begin{figure}[b]
	\includegraphics[width = 1.\linewidth]{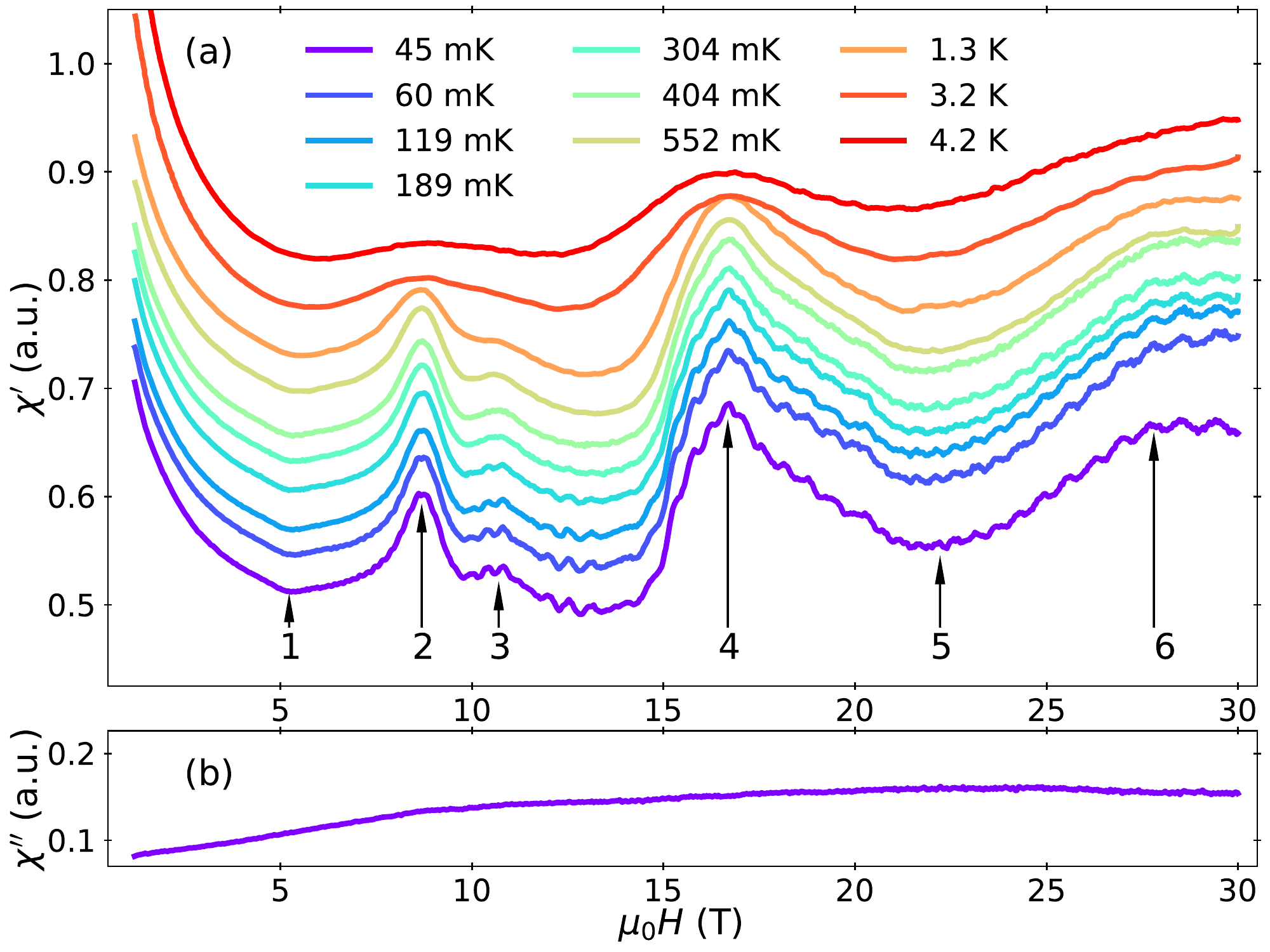}
	\caption{\label{fig:chi_vs_B} 
		(a) The real part of the susceptibility ($\chi^{\prime}$) of sample S1 as a function of magnetic field ($\mu_0H$) for different temperatures; curves are shifted vertically for clarity. $\chi^{\prime}$ shows distinct features in the background shape, which are indicated by the arrows. For the lowest temperatures, one can clearly see quantum oscillations superimposed on the spin susceptibility. (b) The imaginary part of the susceptibility ($\chi^{\prime\prime}$) for the 45~mK curve: $\chi^{\prime\prime}$ is small and there are no clear features. The susceptibility is given in arbitrary units, but the units and the scale are the same in panels (a) and (b). }
\end{figure}

Fig. \ref{fig:chi_vs_B}(a) shows the real part of the magnetic susceptibility $\chi^\prime$ of sample S1 at various temperatures, as a function of the applied magnetic field. Data for sample S2 are similar, and are shown in the Supplementary Information. The curves are shifted vertically relative to each other for clarity. The superconducting transition occurs at $\sim 1$~T \footnote{It is difficult to resolve the upper critical field value very accurately from our susceptibility data because the imaginary susceptibility does not show a clear peak at $H_{c2}$. It seems that the peak is smeared out, possibly because the modulation field affects the pinning of vortices in the mixed state \cite{Gomory}. However, we note that a sharp feature at $H_{c2}$ is also not observed in magnetic torque measurements \cite{Nikitin_thesis}}, but is not shown to allow us to zoom in on the other features in the susceptibility. Fig. \ref{fig:chi_vs_B}(b) shows the imaginary part of the susceptibility $\chi^{\prime\prime}$ at 45 mK. The imaginary signal is small, with no clear features.
$\chi^\prime$ shows distinct features at field values indicated in the figure by arrows labelled $1-6$. The precise locations are $\mu_0H_1= 5.4$~T, $\mu_0H_2= 9$~T, $\mu_0H_3= 11.5$~T, $\mu_0H_4=16$~T, $\mu_0H_5= 22.5$~T, and $\mu_0H_6= 28$~T, which are close to the values previously reported in thermopower and resistivity \cite{Bastien2016, Aoki2011}, and proposed to be field-induced Lifshitz transitions. The demagnetization factors for our samples are negligible \footnote{The demagnetization factor $D_z$ was determined for both S1 and S2, using the expression described by Aharoni\cite{Aharoni}. For S1 we obtain $D_z = 0.084$ and for S2 we obtain $D_z = 0.26$. For S1 this contribution is negligible, which means we can interchange $B$ and $H$ in the analysis. For S2, the difference between $H$ and $B$ is larger. However, the analysis of data from S2 neglecting $D_z$ (Supplementary Information) shows results that are very similar to the results for S1 presented here, so we again conclude that the demagnetization factor is not of high importance.}. We find that  the features 1-6 are significantly broadened with increasing temperature, but most of them remain clearly visible to temperatures well above 1~K.

Superimposed on the spin susceptibility, we also observed well-defined dHvA oscillations in the lowest temperature $\chi^\prime$ curves.
According to the Onsager relation, $F=\hbar A_{\text{ext}}/2\pi e$ \cite{Shoenberg}, the frequency $F$ of the oscillations corresponds to the extremal area $A_{\text{ext}}$ of the Fermi surface in the plane perpendicular to the applied magnetic field. By extracting and tracking the dHvA frequencies we can therefore obtain information about the Fermi surface evolution in an external magnetic field.
 
To follow the field dependence of the frequencies we carried out Fourier transform (FT) analysis over a moving magnetic field window of fixed size in $1/\mu_0 H$. dHvA oscillations are periodic in inverse magnetic field \cite{Shoenberg, Lifshitz1955}, so this approach fixes the number of oscillations of each frequency in a given window, ensuring that the amount of information and resolution in the Fourier transform remains comparable over the whole field range. The field window moves with small steps, such that two consecutive windows will partially overlap.

\begin{figure}[b]
	\includegraphics[width = 1.\linewidth]{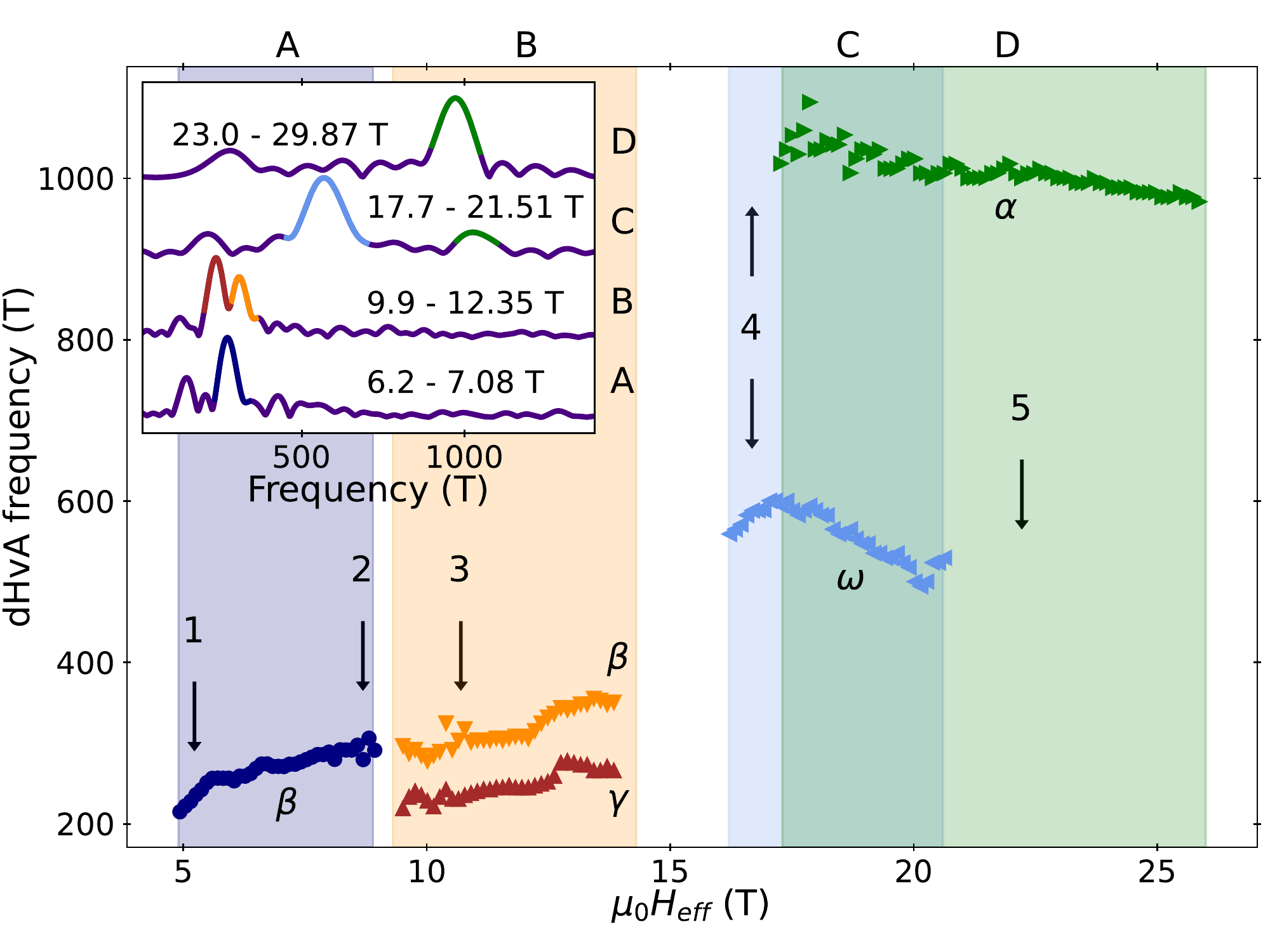}
	\caption{\label{fig:F_vs_B} dHvA frequencies extracted from the $\chi^{\prime}$ vs $\mu_0H$ curve at T = 45~mK, as function of the effective field $\mu_0H_{\text{eff}}$. The data are analyzed over the four regions $A$ to $D$ indicated by the different colors and the letters at the top of the figure. 
	The locations of the field-induced transitions 1-5 are indicated by arrows. The inset shows an example of the FT performed for each field region, $A, B, C, D$, with the frequency peaks we extracted highlighted in different colors. }
\end{figure}

Fig. \ref{fig:F_vs_B} shows the extracted dHvA frequencies from the 45 mK curve as a function of the effective magnetic field, $H_{\text{eff}}=(\frac{1}{2}(\frac{1}{H_L} + \frac{1}{H_H}))^{-1} $, which is the average inverse field for the window over which we performed the Fourier transform. $H_L$ is the low field boundary and $H_H$ is the high field boundary of the analysed window. 

We track each dHvA frequency in the region where it is visible. The series of closely spaced field-induced transitions makes it difficult to track the frequencies continuously with field, so our procedure was to  analyse the data in four different effective field regions $A, B, C, D$, where $A$ covers $4.9 - 8.9$~T, $B$ is $9.3 - 14.3$~T, $C$ is $16.2 - 20.5$~T and $D$ is $17.3 - 26.0$~T, indicated in Fig. \ref{fig:F_vs_B} with different colors. The size of the inverse field window we used for the FTs was constant within each of these regions. The locations of the field-induced transitions 1-5 are also indicated in Fig. \ref{fig:F_vs_B} by arrows. The top left inset shows an example of the FT performed for each different field region, $A, B, C, D$. 

Following the data shown in  Fig. \ref{fig:F_vs_B}: in the lowest field region, $A$, we detected only one frequency ($F_\beta \sim$ 240~T) \footnote{Our results are in agreement with the observed frequencies reported previously from thermopower and magnetoresistance data \cite{Bastien2016, Aoki2011}. We use the same labels, $\alpha$, $\beta$, $\gamma$ and $\omega$, for the observed frequencies.}\footnote{The FTs shown in the inset of Fig. \ref{fig:F_vs_B} also seem to show a small peak at $F \sim 100$~T in regions $A$ and $B$. However, this peak was not reproducible with repeated analysis over different field windows, so we do not consider it to be a real dHvA frequency.}, which grows with increasing magnetic field; in region $B$, we see two frequencies at $F_\gamma = 240$~T and $F_\beta = 310$~T; 
in region $C$, the low frequencies disappear and a high frequency of $F_\omega = 580$~T appears; finally, an even higher frequency of $F_\alpha = 1$~kT is detected in region $D$ and part of region $C$.
$F_\omega$ disappears when crossing the transition at $\mu_0 H_5 = 22.5$~T. 
The detailed magnetic field dependence we have extracted allows us to examine the field-evolution of the different Fermi surface pockets. Combination of these experimental results with calculations of the UCoGe bandstructure 
further allows us to consider various scenarios for Lifshitz transitions. 
In the following sections, we
describe our DFT calculations and then move on to a dicussion of the magnetic field dependence of the Fermi surface.  
\\

\subsection{DFT Calculations}

\begin{figure}[b]
	\includegraphics[width = 1.\linewidth]{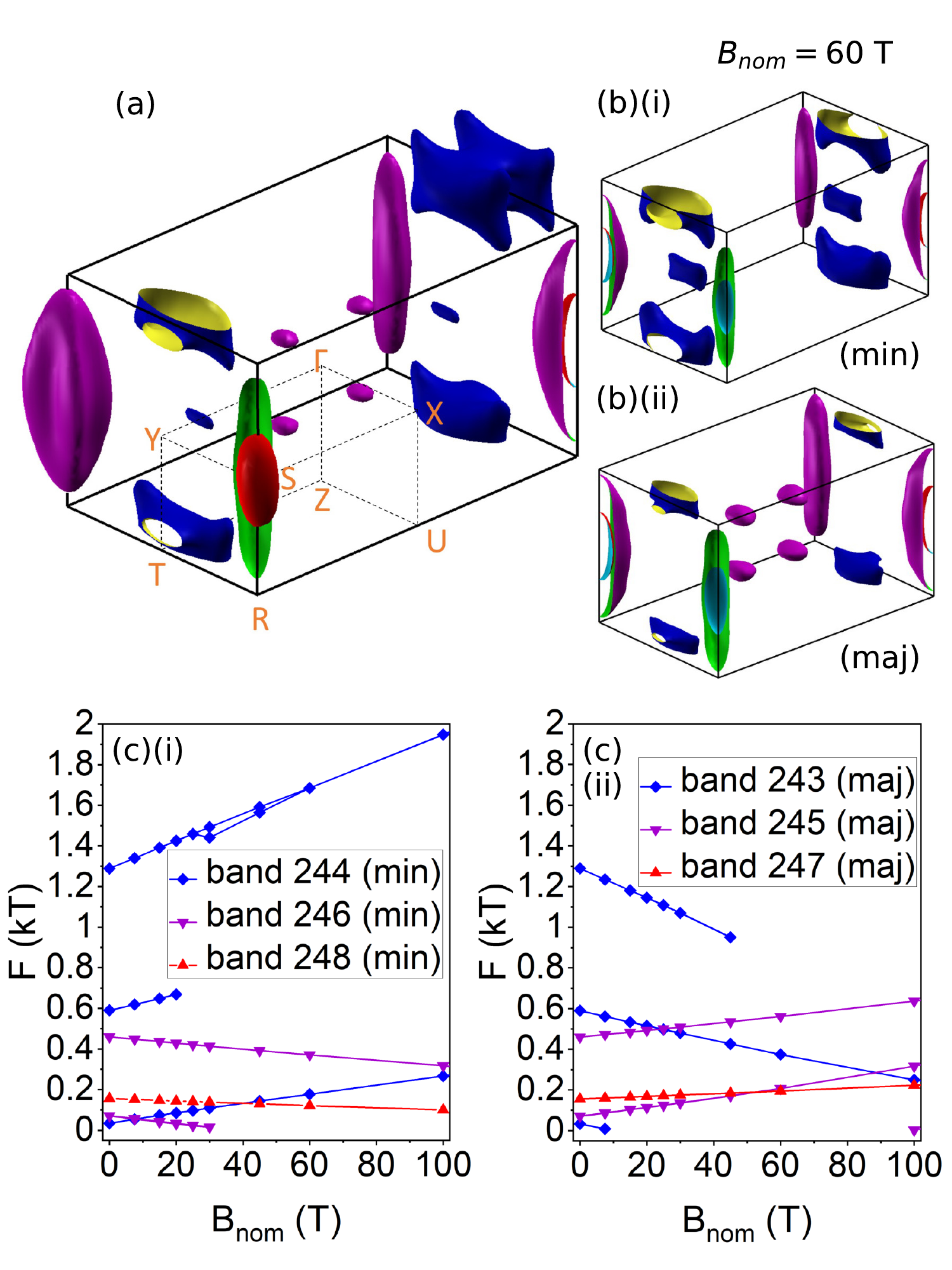}
	\caption{\label{fig:FS_PM_sp} (a) The calculated paramagnetic Fermi surface of UCoGe. Some pockets are extended beyond the first Brillioun zone to show their shape. There are three two-fold degenerate bands that cross the Fermi level. A hole-like band (number 243-244) is indicated in blue/yellow, and two electron-like bands are indicated in purple/green (bands 245-246) and red/light blue (bands 247-248), respectively. (b) shows the effect of a rigid band splitting by a Zeeman term with $B_{\text{nom}} = 60$~T, as described in the text, (b)(i) shows the minority-spin (min) Fermi surface and (b)(ii) the majority-spin (maj) Fermi surface. In panels (c)(i) and (c)(ii) we show the calculated dHvA frequency evolution, $F$, with external magnetic field along the \textit{c}-direction for the minority- and majority-spin carriers, respectively.}
\end{figure}	

We performed DFT calculations with and without spin-polarization. The calculated bandstructures and Fermi surface details are given with further discussion in the Supplementary Information. We found that Fermi surface orbits derived from the calculations performed without spin polarization are in significantly better agreement with the experimental data, so we focus on these results, which we refer to as \lq paramagnetic\rq~calculations. 

The resulting paramagnetic Fermi surface is shown in Fig. \ref{fig:FS_PM_sp}(a). It is very similar to the paramagnetic Fermi surface previously reported by Samsel-Czeka\l a \textit{et al.}\cite{Samsel-Czekala}. Three two-fold degenerate bands cross the Fermi level. The hole band that lies lowest in energy (band 243-244) is indicated in blue/yellow, the two electron bands are shown in purple/green (band 245-246) and red/light blue (band 247-248). At some places in the figure, the pockets are extended beyond the first Brillioun zone to clearly show their shape.

The ferromagnetism of UCoGe is understood to be of weak itinerant nature \cite{Huy, Brando,Moriya1991}. The good agreement between our experimental dHvA frequencies and the paramagnetic DFT calculations is therefore a significant result, as it reinforces this view, implying a Fermi surface that is simply spin-split at zero magnetic field, without marked other effects due to strong polarization.
On this basis, we have modelled the field dependence of the UCoGe Fermi surface by splitting the paramagnetic Fermi surface at zero field, and then increasing the splitting with a Zeeman-like term ($\pm\frac{1}{2}g_0\mu_B B_{\text{nom}}$). $B_{\text{nom}}$ is a nominal magnetic field that captures both the ferromagnetic exchange field and the applied external magnetic field. 
Our susceptibility data suggest weak metamagnetic transitions associated with the observed features 2, 3, 4 and 6, on a monotonically increasing, approximately linear, background (see Supplementary Information, Fig. SI.2), a result supported by previous magnetization data over a similar field range, which shows significant curvature developing only above $\sim 20$~T \cite{Knafo}. Although our model of a linearly spin-split Fermi surface is highly simplified, we therefore believe it provides a reasonable starting point for this first investigation of the magnetic field dependence of the UCoGe bandstructure and Fermi surface. 
Similar models applied to other heavy fermion systems have yielded valuable information 
\cite{McCollam2021, Rourke2008, Daou2006}.

We note that our model gives only a qualitative picture of the effect of an external magnetic field on the Fermi surface, as we cannot relate $B_{\text{nom}}$ directly to the experimentally applied field. The main difficulty is that we do not have information about the $g$-factor, either the zero-field value or any possible magnetic field dependence 
that may arise from complex spin-orbit coupling in this material. 
We have used the free electron value, $g_0 = 2$, for the model, but the real value of $g$ may be considerably different, and the ratio $g/g_0$ is absorbed in $B_{\text{nom}}$.

In Fig. \ref{fig:FS_PM_sp}(b), we show the effect of an applied magnetic field ($B_{\text{nom}} = 60~T$) on the Fermi surface in Fig. \ref{fig:FS_PM_sp}(a): the polarized bands move either up or down in energy, leading to minority-spin (i) and majority-spin (ii) Fermi surfaces. 
Comparing the Fermi surfaces in Fig. \ref{fig:FS_PM_sp}(a) and (b), the most striking differences are that the small ellipsoidal pockets from bands 245-246 have disappeared in the minority Fermi surface, and the middle, cushion-shaped pocket of bands 243-244 has disappeared in the majority Fermi surface. Another notable difference is in the topology of the large pocket of band 243-244 around the T point, which starts to touch the Brillouin zone boundary in the minority Fermi surface. This leads to the hole or tube, visible at the T point in Fig. \ref{fig:FS_PM_sp}(a), being broken open in Fig. \ref{fig:FS_PM_sp}(b)(i).
These changes of the Fermi surface between Fig.s \ref{fig:FS_PM_sp}(a) and (b) clearly represent magnetic field driven Lifshitz transitions.

Fig. \ref{fig:FS_PM_sp}(c) shows the calculated dHvA frequencies for $\mathbf{H}\|c$, and how they evolve with the field $B_{\text{nom}}$. The Fermi surface and quasiparticle orbits were determined at the values of $B_{\text{nom}}$ indicated by the points, and the lines are guides to the eye. Fig. \ref{fig:FS_PM_sp}(c)(i) and (ii) correspond to the field dependence of the minority and majority Fermi surface orbits, respectively.

The disappearances of the 0.071~kT frequency of band 246 in Fig. \ref{fig:FS_PM_sp}(c)(i) and the 0.034~kT frequency of band 243 in  Fig. \ref{fig:FS_PM_sp}(c)(ii) correspond, respectively, to the Lifshitz transitions associated with the disappearances of the ellipsoidal (purple) and cushion-shaped (blue) Fermi surface pockets described above. The disappearances of the 0.59~T frequency (Fig. \ref{fig:FS_PM_sp}(c)(i)) and $\sim$ 1.29~kT frequency (Fig. \ref{fig:FS_PM_sp}(c)(ii)) of band 243-244 are both related to the change of topology of the (blue/yellow) Fermi surface near the T-point.

\subsection{Comparison of experiment with DFT calculations}

We now return to the experimentally determined field dependence of the dHvA frequencies shown in Fig. \ref{fig:F_vs_B}, and use the DFT calculations to help us understand the experimental data.
We first focus on the appearance and disappearance of certain frequencies
as the magnetic field is increased.

In general, the appearance or disappearance of a dHvA frequency can be 
ascribed to a Lishitz transition. However, the presence and amplitude of dHvA oscillations
depends strongly on the applied magnetic field \cite{Lifshitz1955}(see Supplementary Information), and it is also possible for a frequency to \lq appear\rq~with increased 
magnetic field because its amplitude has become large enough to detect above the noise level. 
The appearance of a dHvA frequency, although suggestive, is therefore not 
enough by itself to identify a Lifshitz transition.
An unambiguous Lifshitz transition occurs when a dHvA frequency disappears 
upon increasing the magnetic field.

In the $B_{\text{nom}}$ range shown in Fig. \ref{fig:FS_PM_sp}(c), our DFT calculations give four candidate Lifshitz transitions
associated with the disappearance or change of shape of Fermi surface
pockets. 
In the minority-spin Fermi surface (Fig. \ref{fig:FS_PM_sp}(c)(i)), a frequency from band 244  ($F\sim 0.59$~kT) disappears for  $20 < B_{\text{nom}} < 25$~T  (change in topology of blue/yellow pocket at the T-point). This frequency matches the experimentally observed $F_{\omega}$ (Fig. \ref{fig:F_vs_B}), such
that the disappearance of $F_{\omega}$ could correspond to this predicted Lifshitz transition
on band 244. 
If this is the case, it means that $B_{\text{nom}}$ is close to our experimentally applied magnetic field.
The zero-field exchange splitting is very weak (see Supplementary Information), so the factor accounting for the difference between $B_{\text{nom}}$ and the experimentally applied field comes primarily from the 
ratio of the real $g$-factor of the system and $g_0$.
If we match the range $20\ \text{T} < B_{\text{nom}} < 25$~T 
to the applied magnetic field of $\mu_0H_5 = 22.5$~T, at which we observe the disappearance of $F_\omega$, we can estimate the $g$-factor to be $1.78 < g < 2.22$ for $B \parallel c$. 

The disappearance of pockets from band 246 ($F = 0.071$~kT, small, ellipsoidal, purple Fermi surface), and band 243 ($F = 0.034$~kT, blue, cushion-shaped pocket) also occur in a similar $B_{\text{nom}}$ range, and could correspond to the disappearances of $F_{\beta}$ and $F_{\gamma}$. These small Fermi surface pockets are not those which match best to the dHvA frequencies of $F_\beta$ and $F_\gamma$ (see Supplementary Information), but it is possible that the disappearance of one Fermi surface pocket has an effect on the properties of remaining pockets, particularly those on the same band.
In our calculation, band 243 disappears between $7.5$~T$<B_{\text{nom}}< 15$~T, while band 246 disappears between $30$~T$<B_{\text{nom}}< 45$~T.
In the experiment, $F_{\beta}$ and $F_{\gamma}$ disappear simultaneously, at $\sim 14$~T, so this is a significant discrepancy in the field location of the Lifshitz transition for one of the calculated Fermi surfaces.
We still consider the transition on band 246 to be a candidate, however, as we do not expect our simple model to perfectly capture the behavior of this complicated Fermi surface. For example, our model assumes that all Fermi surface pockets have similar, linear magnetic field dependence, which, as we discuss below, is unlikely to be the case in reality.

The fourth Lifshitz transition predicted for $B_{\text{nom}} < 100$~T, as shown in Fig. \ref{fig:FS_PM_sp}(c), is on the majority-spin Fermi surface, where a frequency ($F\sim 1.29$~kT) from band 243 disappears for $45 < B_{\text{nom}} < 60$~T. If our $B_{\text{nom}}$ is indeed similar to the experimentally applied magnetic field, with a $g$-factor in the range estimated above, this Lifshitz transition on band 243 would be outside of our measured magnetic field range.

Our calculations do not indicate the sudden appearance of any Fermi surface 
pockets over most of the $B_{\text{nom}}$ range shown in Fig. \ref{fig:FS_PM_sp}(c), so we cannot account for the 
experimental appearance of $F_\gamma$ above 9~T or $F_\omega$ and $F_\alpha$ above $\sim 16$~T. As mentioned above, the experimental appearance of the $\gamma$, $\omega$ and $\alpha$ frequencies
may simply be due to the improved resolution given by increasing magnetic field.  
In our calculation at $B_{\text{nom}} = 100$~T, a low frequency appears in band 245, as the
large ellipsoidal pocket around the S point starts to extend to the R corner of the first Brillioun zone and creates a small orbit centred on the R point. This frequency is much lower than any of those observed experimentally.

If we extend our calculations to much higher 
$B_{\text{nom}} > 100$~T, the shape of the Fermi
surface changes significantly, such that several Lifshitz transitions leading to new dHvA frequencies occur on different pockets of the Fermi surface (see Supplementary Information, Fig. 8). These new frequencies are in the correct range to correspond to $F_\gamma$, $F_\omega$ and $F_\alpha$. If we 
consider the experimental disappearance of $F_\omega$, and match $B_{\text{nom}}$ with $\mu_0 H_5$, as we did above, this higher $B_{\text{nom}}$ range would imply 
a minimum $g$-factor of $g = 8.89$ for magnetic field parallel to the $c$-axis.
 
We are not aware of any reports of the $g$-factor in UCoGe. For comparison among other members of the family of U-based superconductors,
Shick {\em et al.} calculate a value of $g = 0.72-0.78$ for \chem{UTe_2}\cite{Shick(2021)}, and in \chem{UGe_2}, Abram \textit{et al.} \cite{Abram2016} 
found their model of the tricritical phase diagram to be in best agreement with experimental data for a value of $g \sim 2$ associated with the $f$ states. The $g$-factor in \chem{URu_2Si_2} was determined from the observation of spin zeros and shows a large anisotropy from $g\|c =2.65\pm0.05$ to $g\|a =0.0\pm0.1$\cite{Altarawneh2012}\cite{Bastien2019}.
Future work to determine detailed $g$-factor information for UCoGe is extremely desirable, as it would allow us to be sure of the quantitative relation between $B_{\text{nom}}$ and the experimentally applied magnetic field. This would, in turn, allow 
a quantitative appraisal of the agreement between the theoretical and experimental results.

As well as the disappearance and appearance of dHvA frequencies, Fig. \ref{fig:F_vs_B} also shows whether the frequencies increase or decrease as a function of magnetic field. Experimentally, however, we measure the {\em back-projection} of the {\em true} frequency $F_{\text{true}}$, which is related to the measured or {\em observed} frequency $F_{\text{obs}}$ according to the following relation\cite{McCollam2013}:
	
\begin{equation}
		F_{\text{obs}}=F_{\text{true}}- B\frac{\partial F_{\text{true}}}{\partial B}
\label{fobs}
\end{equation} 

Because of this relation, we cannot directly relate an increase (decrease) of a frequency as the magnetic field increases to the growth (shrinking) of a Fermi surface pocket.

\begin{figure}[h]
	\includegraphics[width = 1.\linewidth]{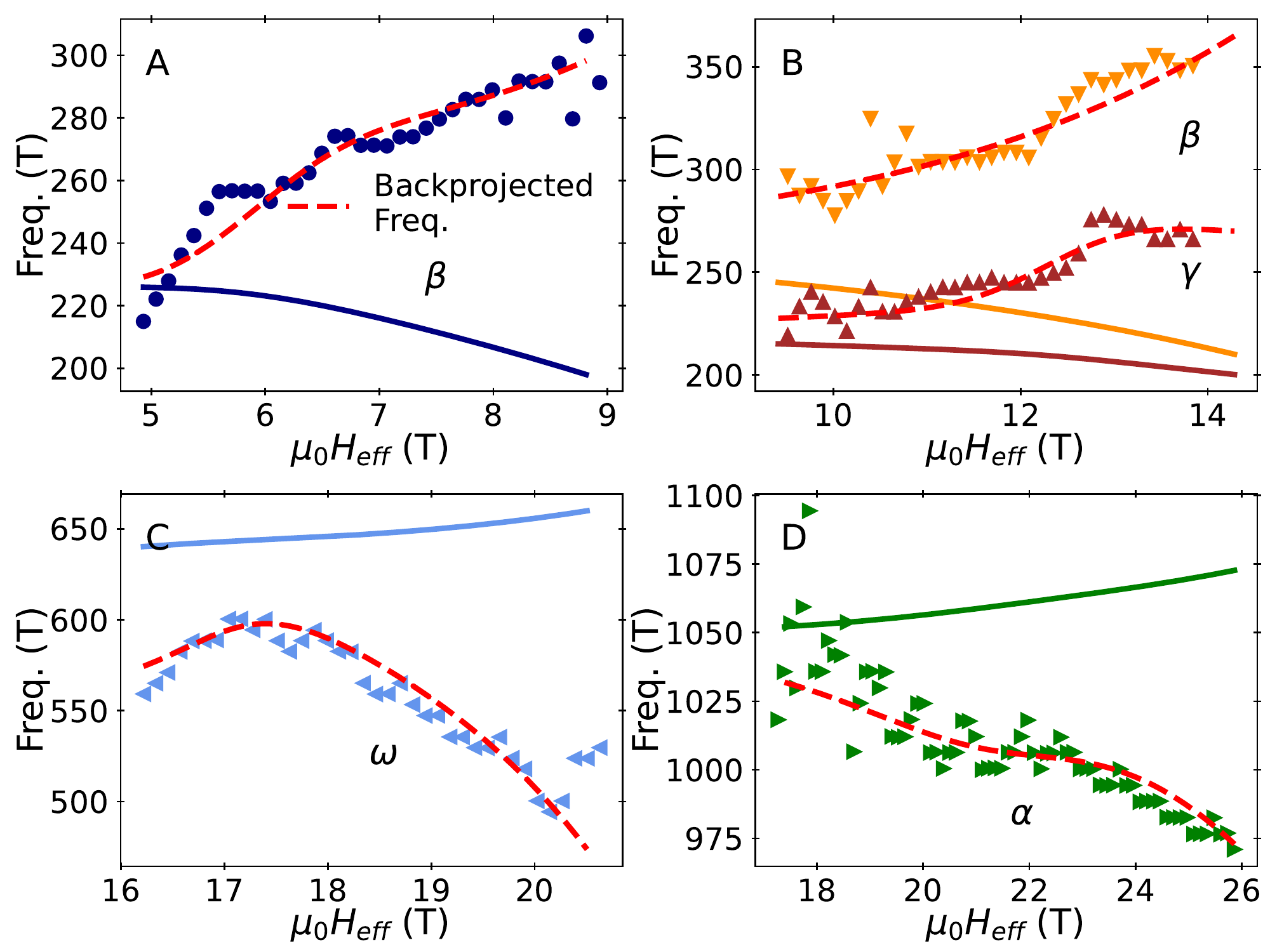}
	\caption{\label{fig:backprojection}The measured magnetic field evolution of the dHvA frequencies ($F_{\text{obs}}$) shown in Fig. \ref{fig:F_vs_B} is shown again for the four different regions $A$ to $D$ as scatter points. A possible true frequency, $F_{\text{true}}$, is depicted in each panel as a solid line, and its backprojection is shown as a dashed red line to demonstrate agreement with the measured data. }
\end{figure}

It is difficult to extract the true frequency from measured dHvA data, but it is worth considering 
the field dependence of the frequencies in more detail to gain some insight into the general behavior.
Fig. \ref{fig:backprojection} shows the different field regions $A-D$ in separate panels. The experimentally measured frequencies, $F_{\text{obs}}$, are indicated as scatter points. A possible true frequency in each case is indicated by a solid line, with the corresponding backprojected frequency shown as a red dashed line to illustrate the agreement with $F_{\text{obs}}$. 
It is particularly striking that the strong curvature of $F_{\omega}$ in region $C$ can be described with a true frequency that is smoothly, and rather weakly, increasing. We note that the true frequencies shown in Fig. \ref{fig:backprojection}
are not definitive, and our observed frequencies can also be reproduced by alternative field dependences of $F_{\text{true}}$
(see Supplementary Information).
However, in matching our measured dHvA data to the paramagnetic Fermi surface calculations, we have tentatively assigned the measured $F_\omega$ frequency (at $\sim 580$~T) to the quasiparticle orbit on the minority hole Fermi surface (band 244), which has a predicted frequency of 590~T undergoes a Lifshitz transition for $20\ \text{T} < B_{\text{nom}} < 25\ \text{T}$. In this case, we expect the Fermi surface to grow, and hence the true frequency to increase with increasing magnetic field, as indicated in Fig. \ref{fig:backprojection}C.

Our observation of field-dependent dHvA frequencies indicates that the
magnetic field evolution of the Fermi surface does not simply depend on
conventional,
linear Zeeman-like spin-splitting. This is clear, for example, from expression (\ref{fobs}),  where a linear field dependence of $F_{\text{true}}$, which has a general shape $( F_{\text{true}}=fB+F_0)$, would result in $F_{\text{obs}}=F_0$, which is the zero-field frequency and is a constant.

Non-linear magnetic field dependence of the Fermi surface can originate in a ferromagnetic material from  Stoner excitations\cite{Ruitenbeek1982}.
Additional contributions to the magnetization, that might give more abrupt Fermi surface changes, can arise from
metamagnetic transitions, which cause sudden increases in the quasiparticle density of states\cite{Tamai2008, Daou2006}.
Another scenario important in some heavy fermion systems is a decrease in the strength of Kondo coupling with increasing magnetic field, causing the $f$-electrons to localize and eventually no longer contribute to the Fermi surface \cite{Friedemann2010, Jiao2015, Mishra2021}. All of these scenarios could, in principle, be relevant to UCoGe, and could manifest differently on different pockets of the Fermi surface. On localization of $f$-electrons, however, the effective mass $m^{*}$ would be expected to decrease with increasing magnetic field \cite{Jiao2019} which is not the case, as can be seen in Fig. SI.3 of the Supplementary Information. Another possible cause of anomalous magnetization in UCoGe is the field-dependent cobalt moment, which is neglegible at zero field, and starts to grow non-uniformly when magnetic field is applied \cite{Prokes, Taupin, Butchers, Bay2014}.

We now turn to the additional information we can extract from the quantum oscillations. We measured the oscillations at different temperatures, and can determine the effective mass $m^{*}$ of the quasiparticles from the temperature dependence of the oscillation amplitudes \cite{Shoenberg}. 
We obtained masses ranging from $9.7\pm1.4$~$m_e$ (where $m_e$ is the bare electron mass) for $F_\beta$ in field region $A$, to $13.2\pm1.0$~$m_e$ for $F_\alpha$ in region $D$, without any strong magnetic field dependence. The full set of results for $m^*$ is shown in the Supplementary Information. 

We also tracked the amplitude of the dHvA oscillations as a function of magnetic field, and used this to extract the Dingle temperatures, which give a measure of the scattering on the Fermi surface \cite{Shoenberg}. The Dingle temperatures range from $0.30\pm0.01$~K for $F_\beta$ in field region $A$ to $0.85 \pm0.01$~K for $F_\alpha$ in region $D$. These values indicate similar scattering on all Fermi surface pockets and relatively long quasiparticle mean free paths of a few tens of nanometers, as we would expect from samples with high RRR's. The full results and discussion of the Dingle factors are also given in the Supplementary Information. 

In the foregoing discussion, we considered the field dependence of the experimentally observed dHvA frequencies and quasiparticle properties in relation to the predicted field dependence of the Fermi surface extracted from our calculations. 
We now want to explicitly relate the field dependence of the dHvA frequencies to the features in $\chi^\prime$ (numbered $1-6$ in Fig. \ref{fig:chi_vs_B}).

Features $\mu_0H_1$ and $\mu_0H_6$ in $\chi^\prime$ are at the edges of our resolution and field range. We cannot resolve quantum oscillations in our data below $\mu_0H_1\sim 5.4$~T, and the available magnetic field was not high enough for us to to reliably extract dHvA frequencies above $\mu_0H_6\sim 28$~T, so we are unable to comment on the field dependence of the Fermi surface across these two features. 

Fig. \ref{fig:F_vs_B} shows that $F_{\gamma}$ first appears in our data at $\mu_0H_2 \sim 9$~T. 
However, evidence for a Lifshitz transition at $\mu_0H_2$ is weak.   
Previous measurements of quantum oscillations in the transverse magnetoresistance identify $F_{\gamma}$ both above and below $\mu_0H_2$ \cite{Bastien2016}, and this, combined with the smooth field dependence of $F_{\beta}$ that we show in Fig. \ref{fig:F_vs_B}, suggests a normal field-evolution of the Fermi surface through this region.
The origin of the large peak in $\chi^\prime$ at $\mu_0H_2$ therefore remains an open question.
Earlier work suggested that a feature near 9~T$||c$  in several properties of UCoGe, including magnetic susceptibility, was due to a ferro- to ferrimagnetic transition, driven by a field-induced moment on the Co site, aligned antiparallel to the U moment \cite{Prokes, Steven, Butchers, Bay2014}. These suggestions were apparently ruled out in reference \cite{Taupin}, where the authors provided evidence that the U and 
Co moments are parallel, and proposed that a Lifshitz transition is responsible for the observed amomalous behaviours near 9~T.
Our results, showing weak metamagnetism and little evidence for a Lifshitz transition at $\mu_0H_2 \sim 9$~T, have re-opened at least some aspects of this discussion, and prompt further work to understand the development and influence of magnetization associated with the Co $3d$ orbitals. Inclusion of the field-dependent Co moment in bandstructure calculations could be informative in this respect.

The peak in $\chi^\prime$ at $\mu_0H_3\sim 11.5$~T is well-defined up to temperatures of 1.3~K, and there is no evidence from our dHvA data of a Fermi surface change at this field. Apart from its significantly smaller amplitude, the feature at $\mu_0H_3$ is therefore similar to that at $\mu_0H_2$, and further investigation of the contributions to magnetization in UCoGe is needed to understand its origin.

Our experimental data suggest that the features in $\chi^\prime$ at both $\mu_0H_4$ and $\mu_0H_5$ are associated with Lifshitz transitions. This is supported by our model of the field dependence of the Fermi surface. We note however, that the shapes of the features at $\mu_0H_4$ and $\mu_0H_5$ are very different: a large peak at $\mu_0H_4$, corresponding to a clear step in the magnetization (Supplementary Information, Fig. SI.2), compared to a broad minimum in $\chi^\prime$ that translates to a weak inflection point in the magnetization at $\mu_0H_5$.
In our interpretation, both Lifshitz transitions involve Fermi surfaces of a single spin orientation, but the Lifshitz transition at $\mu_0H_5$ is a change of topology of the largest Fermi surface pocket, whereas the transition or transitions at $\mu_0H_4$ are the disappearances of small Fermi surface pockets. 
It is interesting to consider if the larger feature at $\mu_0H_4$ reflects 
a direct effect on the itinerant moment when a spin-polarized Fermi surface pocket disappears.
It would be useful to study in more detail how different types of Lifshitz transition influence the magnetic susceptibility and wider properties of the system, as this can shed light on the relative contributions of specific bands to the overall magnetization.

\section{Conclusion}
In summary, we measured the magnetic susceptibility of high quality single crystals of UCoGe at low temperatures and in high magnetic fields.
For magnetic field parallel to the crystal $c$-axis, we observe a series of clearly-defined features in $\chi^{\prime}$, as well as de Haas-van Alphen oscillations with multiple, strongly field-dependent frequencies, which suggest that field-induced Lifshitz transitions occur.

We combined our experimental results with DFT calculations of the 
UCoGe bandstructure,  specifically to study the shape and field-evolution 
of the Fermi surface. 
We find that calculations of a paramagnetic Fermi surface better match our
experimental dHvA results than a fully spin-polarized Fermi surface, and 
argue that this is in keeping with the weak itinerant ferromagnetic 
moment in this material.
On this basis, we used a simple model of a spin-split paramagnet to estimate the magnetic field dependence of the Fermi surface.

Through comparison of the measured and calculated magnetic field 
dependence of the dHvA frequencies, we identified candidate Lifshitz transitions that could account for two of the experimentally observed features in $\chi^\prime$: disappearance of the $\beta$ and $\gamma$ frequencies in the region of $\mu_0H_4$,
and disappearance of the $\omega$ frequency at $\mu_0H_5$. 
We show how these Lifshitz transitions correspond to changes in topology or disappearance of specific Fermi surface pockets.
To confirm that these specific Lifshitz transitions are responsible for the experimental observations would require information about the $g$-factor of the system, which is needed to make a definite relation between the nominal magnetic field in our model and the actual applied field in the experiments. 
The candidate Lifshitz transition we discuss for $F_{\omega}$ would fix $1.78 < g < 2.22$. 
However, further Lifshitz transitions, which could also be in keeping with the experimental data, occur at much higher $B_{\text{nom}}$, and would imply a $g$-factor above 8.89. 
This high value of $g$ seems unrealistic, and calculations or additional experiments aimed at determining $g$ would therefore be highly valuable.

Through further detailed analysis of the dHvA oscillations, we determined the field dependence of the quasiparticle effective masses, and also extracted Dingle temperatures to gauge the scattering on different pockets of the Fermi surface. 
The effective masses are moderate for a heavy fermion material and are similar on all Fermi surfaces, of order $10~m_e$, in keeping with previous results \cite{Aoki2011,Bastien2016}. 
The Dingle temperatures and estimated quasiparticle mean free paths, in the range 20-50 nm, reflect the relatively low scattering in our high quality single crystal samples.
Importantly, our results for the effective masses and the Dingle factors show that there is only a weak magnetic field dependence of the quasiparticle properties, as well as a weak variation of properties between Fermi surface pockets, even as the system is tuned through significant Lifshitz transitions.

Understanding the Fermi surface and intinerant quasiparticles in the U-based ferromagnetic or nearly ferromagnetic superconductors is a key challenge in the field of strongly correlated electron systems, as they play a crucial connecting role 
between superconductivity and magnetism.
Our results on UCoGe give the first detailed information about the magnetic field dependence of its Fermi surface properties.
The insight we have provided into the nature of the Lifshitz transitions in UCoGe helps us to understand the relation of the Fermi surface to the magnetic response of the system, and identifies important questions for further study.

Efforts to develop an improved model of the spin-splitting and magnetic field dependence of the UCoGe Fermi surface are strongly motivated by the present results.
Incorporating the real, non-linear field dependence of the magnetization, as a reflection of the field-dependent exchange splitting, would be one way to do this \cite{McCollam2021, Lonzarich1974, Yelland2007}. It would require more precise magnetization data over a wide field and temperature range than are currently available, but could remove much of the approximation in identifying the field-location of the Lifshitz transitions in the model.  
Another approach is to determine the field-dependent bandstructure with advanced theoretical techniques, such as the renormalized band method \cite{Zwicknagl1992, Zwicknagl2016, Pourret2019}.
Combining an improved model with an investigation into the contribution of cobalt to the 
bandstructure and magnetization as the magnetic field is tuned would be 
particularly informative.

The results we present here are also relevant in a wider context, as there is an increasingly extensive tableau of strongly correlated and topological materials, spanning from graphene to iron-based superconductors and archetypal heavy fermion systems,
\cite{Jayaraman2021, Joshua2012,Wu2023, Coldea2019, Quader2014, Ptok2017,Pourret2019, Daou2006, Pfau2013, Pfau2017,	McCollam2021, Volovik2017}
in which Lifshitz transitions are found to have significant influence on behavior and properties. This creates a pressing need to explore and understand Lifshitz transitions in diverse settings.

\begin{acknowledgments}
We would like to thank S. R. Julian and G. Zwicknagl for valuable discussions. This work was supported by HFML-RU/NWO-I, a member of the European Magnetic Field Laboratory (EMFL) and the French National Agency for Research ANR within the project FRESCO No. ANR-20-CE30-0020 and FETTOM No. ANR-19- CE30-0037. 
\end{acknowledgments}

\end{document}

\title{Supplementary Information: Fermi Surface and Lifshitz Transitions of a Ferromagnetic Superconductor under External Magnetic Fields}
	
\author{R. Leenen}
\affiliation{High Field Magnet Laboratory (HFML-EMFL), Radboud
University, Toernooiveld 7, 6525 ED Nijmegen, The Netherlands.}
\affiliation{Radboud University, Institute for Molecules and Materials, Heyendaalseweg 135, 6525 AJ Nijmegen, The Netherlands.}	
\author{D. Aoki}
\affiliation{Institute for Materials Research, Tohoku University, Ibaraki 311-1313, Japan}
\author{G. Knebel} 
\affiliation{Univ. Grenoble Alpes, CEA, Grenoble INP, IRIG, PHELIQS, F-38000 Grenoble, France}
\author{A. Pourret} 
\affiliation{Univ. Grenoble Alpes, CEA, Grenoble INP, IRIG, PHELIQS, F-38000 Grenoble, France}
\author{A. McCollam}
\affiliation{High Field Magnet Laboratory (HFML-EMFL), Radboud
University, Toernooiveld 7, 6525 ED Nijmegen, The Netherlands.}
\affiliation{Radboud University, Institute for Molecules and Materials, Heyendaalseweg 135, 6525 AJ Nijmegen, The Netherlands.}

\maketitle

\section{More on the experimental results}
\subsection{Comparison of results from samples S1 and S2}

\begin{figure}[h]
	\includegraphics[width = 1.\linewidth]{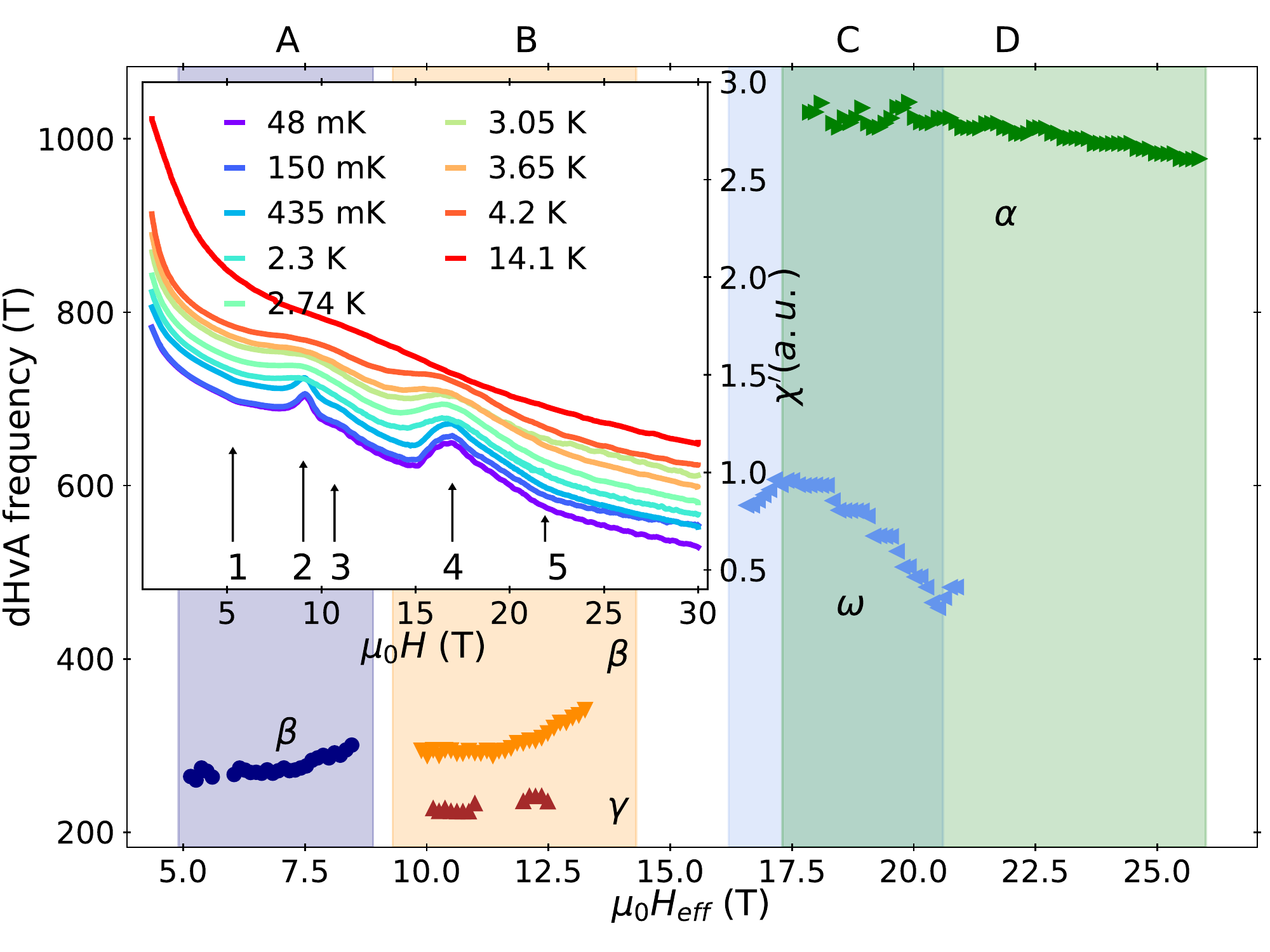}
	\caption{\label{fig:RRR=40_results} dHvA frequency vs $\mu_0H_{\text{eff}}$ analyzed from the $\chi^{\prime}$ curve of S2 taken at 48~mK. The regions $A - D$ are the same regions indicated in the main paper for S1. The inset shows the data of $\chi^{\prime}$ vs $\mu_0H$ for different temperatures, with the features observed in $\chi^{\prime}$ indicated by the arrows.}
\end{figure}

In the main paper we present the frequency dependence of the dHvA oscillations observed in UCoGe sample S1 with increasing magnetic field (Fig. 2). We performed the same analysis on the $\chi^{\prime}$ vs $\mu_0H$ data of sample S2 taken at $T=48$~mK. The results are presented in Fig. \ref{fig:RRR=40_results}.  
The figure shows the frequency evolution of all the detected oscillations with applied magnetic field. The regions $A - D$ are the same field regions indicated in the main paper. We can see that the frequency evolution is very similar to the results obtained for S1, although $F_{\gamma}$ cannot be resolved in the entire region $B$. The inset shows the $\chi^{\prime}$ vs $\mu_0H$ curves at different temperatures. The distinct features in the $\chi^{\prime}$ curve are indicated by arrows. 

\begin{figure}[h]
	\includegraphics[width = 1.\linewidth]{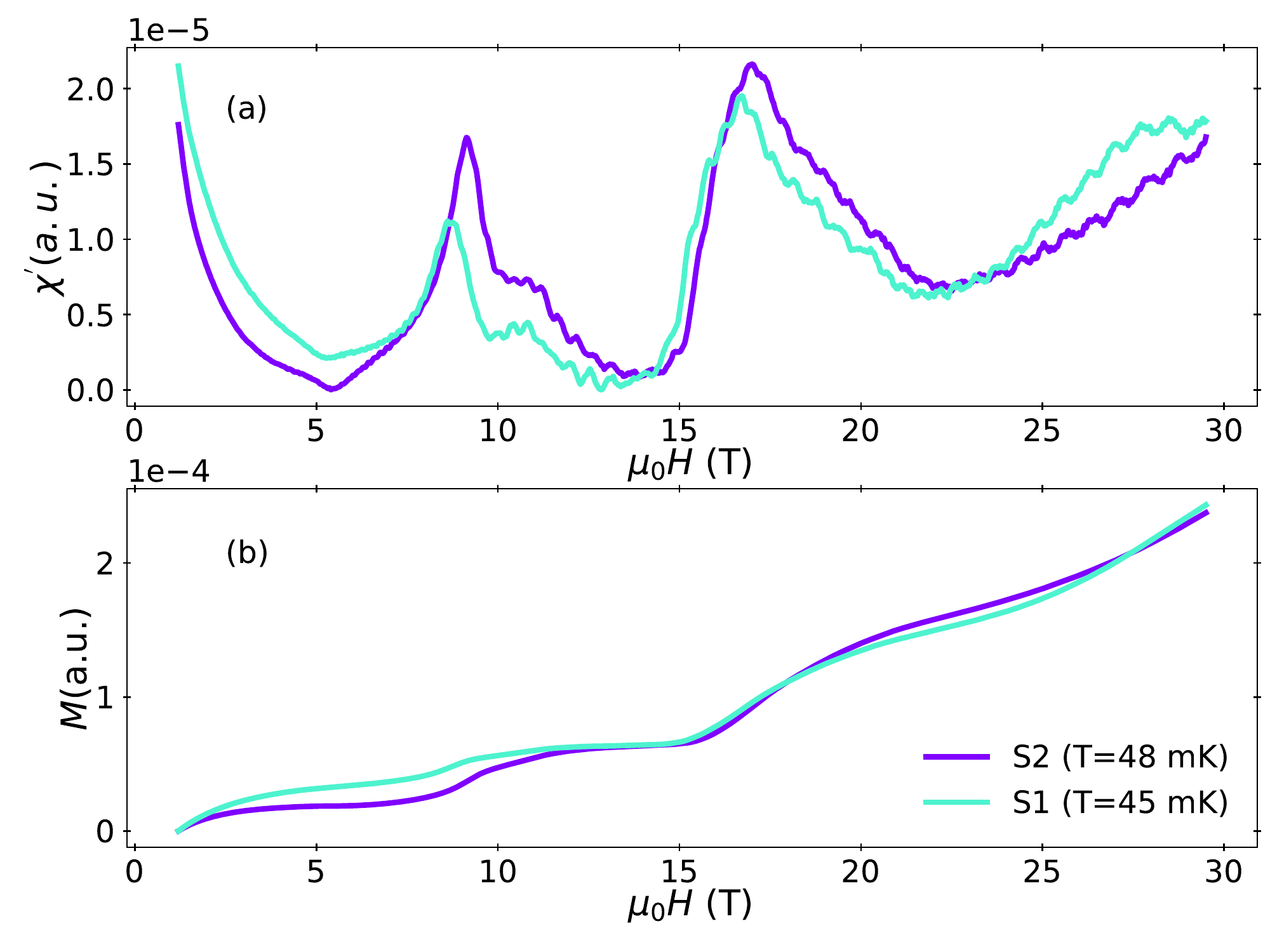}
	\caption{\label{fig:integrated} (a) Comparison of the $\chi^{\prime}$ signal of both S1 and S2. A linear background is subtracted of the $\chi^{\prime}$ curve of S2 to make it follow a similar trend as the $\chi^{\prime}$ curve of S1. (b) The integrated susceptibility $M = \mu_0\int\chi^{\prime}dH$ (as shown in (a)) for both samples. }
\end{figure}

Samples S1 and S2 have different dimensions. We used different sets of pick-up coils to measure S1 and S2, to maximize the filling factor in each case. The  overall susceptibility signal we measure depends partly on the sample and partly on the background from the pick-up coils, which may not be perfectly balanced over the entire magnetic field range. It is therefore not surprising that the background slope is different for S1 and S2. If we make a correction to the S2 data to give a similar $\chi^{\prime}$ background to S1 (see Fig. \ref{fig:integrated} (a)) we can compare them more easily.

The remaining differences between the overall shapes of the curves from the two samples are likely because of a slight difference in orientation with respect to the magnetic field. UCoGe is highly anisotropic, and sensitive to changes in angle of fractions of a degree away from the \textit{c}-axis \cite{Huy2008, Knafo}. This would explain, for example,  why the locations of the features are not exactly the same in both curves; the S2 features are shifted to a slightly higher field. 
A slight misalignment might also explain why the $\mu_0H_6$ feature of S1 (see Fig. 1 in the main paper) is not visible in S2, because it is beyond the magnetic field we can reach with our set-up. The accuracy of alignment in the experimental set-up was approximately 1 degree.

To consider the magnetization of the sample, we integrated the susceptibility signals of both S1 and S2. The results are shown in Fig. \ref{fig:integrated} (b).
The peaks in susceptibility translate to steps in the magnetization, as we would expect. It should be noted that the shape of the integrated curve is very sensitive to the overall background of the $\chi^{\prime}$ vs $\mu_0 H$ curve, and this background depends on the pick-up coils used. Because of this, we can interpret the general behavior of the magnetization curve, but do not have information about its exact shape and magnitude.

In a weak itinerant ferromagnet with Stoner-like excitations, the magnetization is expected to reflect the exchange splitting of the bands \cite{Lonzarich1974, Yelland2007}. A more sophisticated model of the field dependence and spin-splitting of the Fermi surface could therefore be developed by modelling the field-dependent exchange splitting based on the real, non-linear magnetization of the material. However, uncertainty about the background signal of the magnetisation from the integrated susceptibility that we show in Fig. \ref{fig:integrated}, make it unsuitable for incorporation into the model in this way. Precise and well-calibrated magnetization data, over a high field/low temperature range, would be required.

\subsection{Procedure of analysis of experimental data}
To extract information about the dHvA frequencies from the experimental data, we carried out Fourier transform (FT) analysis. The background signal of $\chi^{\prime}$ is not trivial, as can be seen in Fig. 1 in the main paper, so for each field region we look at, we subtract a polynomial background. 
For S1, we first subtracted a smoothed high temperature curve, which did not contain oscillations, to get rid of a large part of the background. After the background subtraction, we applied a running average to smooth the experimental data and we performed FTs over a moving magnetic field window of fixed size in $1/\mu_0 H$.

We analyzed dHvA oscillations in the different field regions $A-D$, as indicated in the main paper, with a fixed inverse magnetic field range in each region. The field window moves with small steps, such that two consecutive windows will partially overlap. We do not use a window function for the FT and we apply a Bessel function correction to obtain the correct amplitude \cite{Shoenberg}. The frequencies we obtain from the FT, and the corresponding amplitudes, are coupled to the effective magnetic field, $H_{\text{eff}}=(\frac{1}{2}(\frac{1}{H_L} + \frac{1}{H_H}))^{-1} $, where $H_L$ is the low field boundary and $H_H$ is the high field boundary of the analyzed window. 

\subsection{Quasiparticle effective masses and Dingle temperatures extracted from the experimental data of S1}
\begin{figure}[h]
	\includegraphics[width = 1.\linewidth]{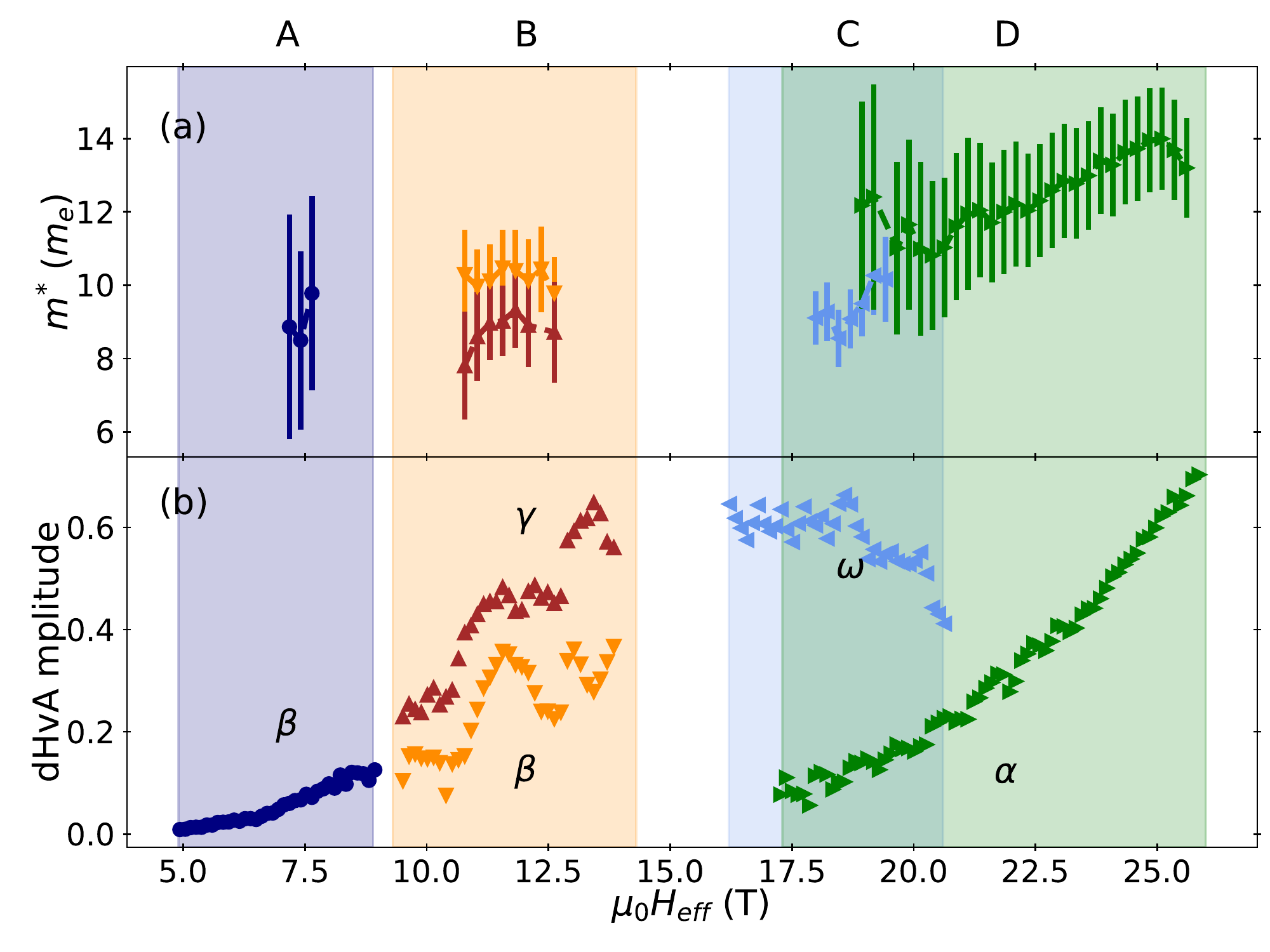}
	\caption{Further analysis of S1 data where regions $A-D$ are the same as in the main paper: (a) Shows the field dependence of the effective mass $m^*$. The error bars are quite large due to the small magnetic field regions that are used to analyze the dHvA oscillations, and due to the temperature uncertainty at higher fields. (b) shows the field dependence of the amplitudes of the oscillations, taken as the amplitudes of the peaks in the FT. }\label{fig:m_and_A_vs_B} 
\end{figure}

In Fig. 2 of the main paper, we show the evolution of the dHvA frequencies as a function of applied magnetic field extracted from the 45~mK $\chi^{\prime}$ curve for S1. In a similar way, we can follow the amplitude evolution of the dHvA signals by extracting the field dependence of the amplitude of the peaks in the FT spectra. The result of this analysis is shown in Fig. \ref{fig:m_and_A_vs_B}(b).

By fitting the temperature dependence of the amplitude at a given (effective) magnetic field with the temperature damping term in the Lifshitz-Kosevich formula \cite{Shoenberg}, 
\begin{equation}\label{eq:R_T}
R_T=\frac{x}{\sinh(x)}\ \ \ \ \ \ \ \text{where}\ \ \ \ \ \ x=14.69\frac{m^{*}T}{B}
\end{equation}
we can extract the quasiparticle effective mass ($m^*$). By repeating this analysis of temperature dependence for the moving window FTs described above and in the main paper, we find the field dependence of $m^*$, which is shown in Fig. \ref{fig:m_and_A_vs_B} (a).
Because the magnetic field regions over which we perform the FTs are quite small (restricted by the presence of the features in $\chi^\prime$), the FT resolution is limited and the $m^*$ we determine is subject to a large uncertainty, indicated in Fig. \ref{fig:m_and_A_vs_B} (a).
An uncertainty in temperature during the highest field parts of our measurements
\footnote{It is a well-known problem that liquid $^4$He in the bath cryostat begins to levitate in very high magnetic fields. In slow sweeps of the magnetic field, this can lead to a ring or bubble of gas forming in the cryostat near the field centre, which causes a heat leak to the sample region of the dilution refrigerator. We are able to account for this to some degree, but it leads to additional uncertainty in the temperature at the sample, in our case above $\sim 20$~T.}
is also incorporated into the errors bars shown in Fig. \ref{fig:m_and_A_vs_B} (a). The large error bars mean that we can only identify general trends, such as the small growth of $m^*$ with increasing field in field region $D$, rather than precise behavior. Some representative values of the effective mass are shown in table \ref{tab:effective_mass}. 
For the values in table \ref{tab:effective_mass}, we analyzed the dHvA oscillations over a larger magnetic field region than was used for the data in figure \ref{fig:m_and_A_vs_B} (a) (except for region $D$) to reduce the error in the mass estimate.
\begin{table}[h]
	\begin{tabular}{l|l|l}
		$\mu_0H_{\text{eff}}$ region & $\mu_0H$ range (T)&	$m^*$ ($m_e$) \\
		\hline
		\hline
		$A$ & 6.9-8.3 & $9.673 \pm 1.436$ ($F_{\beta}$)\\
		\hline
		$B$ & 9.3 -13.5 & $10.571 \pm 0.525$ ($F_{\beta})$\\
		& & $9.094 \pm 0.439$ ($F_{\gamma}$)\\	
		\hline
		$C$ & 16.8-20.8 & $9.818 \pm 0.315$ ($F_{\omega}$)\\
		& & $12.585 \pm 1.227$ ($F_{\alpha}$)\\
		\hline
		$D$ & 24.2-29.9 & $13.173 \pm 0.974$ ($F_{\alpha}$)\\
	\end{tabular}
	\caption{The table shows typical values of the effective mass for each of the field regions $A-D$. The effective mass is determined by fitting the temperature damping term of Lifshitz-Kosevich formula, expression \ref{eq:R_T}, in each region. The corresponding frequency label is shown in brackets. The exact field region we analyzed over is also shown in the table. We note that these field regions are larger than the field region used to produce figure \ref{fig:m_and_A_vs_B} (except for region $D$), and therefore the error in these estimates is generally smaller than those shown in figure \ref{fig:m_and_A_vs_B}. }  
\label{tab:effective_mass}
\end{table}

Turning to the field dependence of the amplitude of the dHvA signal shown in Fig. \ref{fig:m_and_A_vs_B} (b), we can estimate the Dingle temperature, which gives a measure of the quasiparticle scattering in the material.
The Dingle temperature $T_D$ is given by 
\begin{equation}\label{eq:T_D}
	T_D = \frac{\hbar m_b}{2\pi^2k_B\tau m^*}
\end{equation}
where $m_b$ is the band mass, $m^*$ is the effective mass of the quasiparticles and $\tau$ is the scattering time.
We extracted a Dingle temperature from our magnetic susceptibility data by fitting the following relation for the amplitude $\cal{A}$ as function of magnetic field $B$ \cite{Shoenberg}:
\begin{equation}\label{eq:Dingle}
	{\cal{A}} = s *F* B^{-5/2}\frac{ \exp{(-2\pi^2 k_B m^* T_D/\hbar e B)}}{\sinh(2\pi^2 k_B m^* T/\hbar e B)}
\end{equation}
where $T$ is the temperature and $F$ is the dHvA frequency. $s$ contains the spin damping term and the Fermi surface curvature term, both of which we assume to be constant. 
Because of the uncertainty in the effective mass and temperature values, we cannot predict the Dingle temperature very accurately. 
In region $C$ the dHvA amplitude decreases with increasing field, which is anomalous behavior. The fit to expression \ref{eq:Dingle} is not reliable in this region, so we only present the results for regions $A,B$ and $D$.
To simplify the fitting procedure we took a fixed effective mass, corresponding to a representative mass of the specific field region (see table \ref{tab:effective_mass}), and did not take the weak field dependence of the mass into account. The fitting of $T_D$ was done both with and without including the measured field dependence of the dHvA frequency. The difference between the $T_D$ values obtained upon including the field dependence of the dHvA frequency or keeping the frequency constant is small for regions $A$ and $D$, but more substantial for region $B$. The results for each region are shown in table \ref{tab:Dingle}.

\begin{table}[h]
	\begin{tabular}{l|l|l|l|l|l}
		\vtop{\hbox{\strut $\mu_0H_{\text{eff}}$ }\hbox{\strut region}} & $F$ & 		\vtop{\hbox{\strut Field }\hbox{\strut dependent $F$?}} &	$T_D$ (K)& $r^{2}$ & \vtop{\hbox{\strut $\lambda$ }\hbox{\strut (nm)}} \\
		\hline
		\hline
		A & $F_{\beta}$ &yes & $0.304 \pm 0.011$& 0.977 & 43 \\
		&  & no & $0.322 \pm 0.011$& 0.974 & 41\\
		\hline	
		B & $F_{\gamma}$ & yes &$0.256\pm0.013 $& 0.886& 45\\
		&  & no & $0.299 \pm 0.013$ & 0.894& 39 \\
		& $F_{\beta}$ & yes &$0.285\pm 0.038$ &0.520& 53\\
		&  & no &$0.341 \pm 0.013$& 0.602& 44\\
		\hline
		D & $F_{\alpha}$ &yes & $0.852 \pm 0.009$ & 0.993& 22 \\
		& & no &$0.830 \pm 0.008$ & 0.994 & 23\\
	\end{tabular}
	\caption{The Dingle temperature $T_D$ is determined for the different field ranges $A$, $B$ and $D$, by using expression (\ref{eq:Dingle}). $r^2$ gives a measure of the quality of the fit. The amplitude evolution of the dHvA frequency in region $C$ was anomalous and could not be fitted with expression (\ref{eq:Dingle}).  The fitting procedure was performed with and without taking the field dependence of $F$ into account. $m^{*}$ was taken to be constant and a representative value for the region was used (see table \ref{tab:effective_mass}). The mean free path $\lambda$, estimated from the Dingle temperature using expression \ref{T_D:mfp}, is given in the rightmost column.} \label{tab:Dingle}
\end{table}

For regions $A$ and $D$ the amplitude behaves in a typical way, increasing roughly exponentially with field, and we get a good fit to the $T_D$ expression.
In region $B$ the amplitude does not grow smoothly with increasing field, due to the beating pattern caused by the closely spaced frequencies $F_{\beta}$ and $F_{\gamma}$. This makes it more difficult to accurately determine the Dingle temperature, and the fit to expression \ref{eq:Dingle} in this region is therefore not good. The quality of the fit to $T_D$ is indicated by the $r^2$ value from the fitting procedure, which is also shown in 
table \ref{tab:Dingle}.

The Dingle temperature can be recast in terms of the Fermi velocity $v_F$ and quasiparticle mean free path $\lambda$ on the Fermi surface:
\begin{equation}
T_D = \left(\frac{\hbar}{2\pi k_B}\right) \frac{v_F}{\lambda}.
\label{T_D:mfp}
\end{equation}
Because the dHvA frequency is related to the area of the quasiparticle orbit on the Fermi surface $A_{\text{ext}}$  
via the Onsager relation,
$F = \hbar A_{\text{ext}} / (2 \pi e)$ 
we can use the measured dHvA frequencies and effective masses to 
calculate approximate values of the mean free 
path on each measured quasiparticle orbit.
We assume circular Fermi surface orbits to 
estimate 
$A_{\text{ext}} = \pi k_F^2$ 
and hence $v_F = \hbar k_F/m^*$.  
The values of $\lambda$ then obtained from (\ref{T_D:mfp}) are given in the 
rightmost column of table \ref{tab:Dingle}
\footnote{The quasiparticle orbits are not circular, but from the 
dHvA experiments alone we have no information about their actual shape. 
The approximation of a circular orbit is reasonable for the purpose of 
estimating $\lambda$ in this context because both $m^*$ and $v_F$ are 
the average of their values around a given orbit.}.

As noted above, the anomalous behavior of the dHvA amplitude in region $C$ makes the fit to expression \ref{eq:Dingle} unreliable.
We note that a field-dependent spin damping or curvature term (in equation \ref{eq:Dingle}) can introduce extra amplitude  damping; we took these terms to be constant, but this might not be a correct assumption in region $C$. 
For example, the curvature term can change if the Fermi surface changes shape, which might be the case at or near Lifshitz transitions. 
Field dependence of the spin damping term can arise if spin-up and spin-down amplitudes are not equal, if the phase difference between spin-up and spin-down frequencies is field dependent, or if the effective mass or $g$-factor is field dependent. 

Other examples of heavy fermion systems which experience damping of their dHvA amplitudes after a field induced transition are \chem{CeIrIn_5}, where selective damping of the heavier pockets is suggested to occur above a field induced transition\cite{Capan}, and URhGe, where the decreasing amplitude near a Lifshitz transition is appointed to a shrinking Fermi surface \cite{Yelland}.

\subsection{Backprojection with alternative true frequency}
\begin{figure}[h]
	\includegraphics[width = 1.\linewidth]{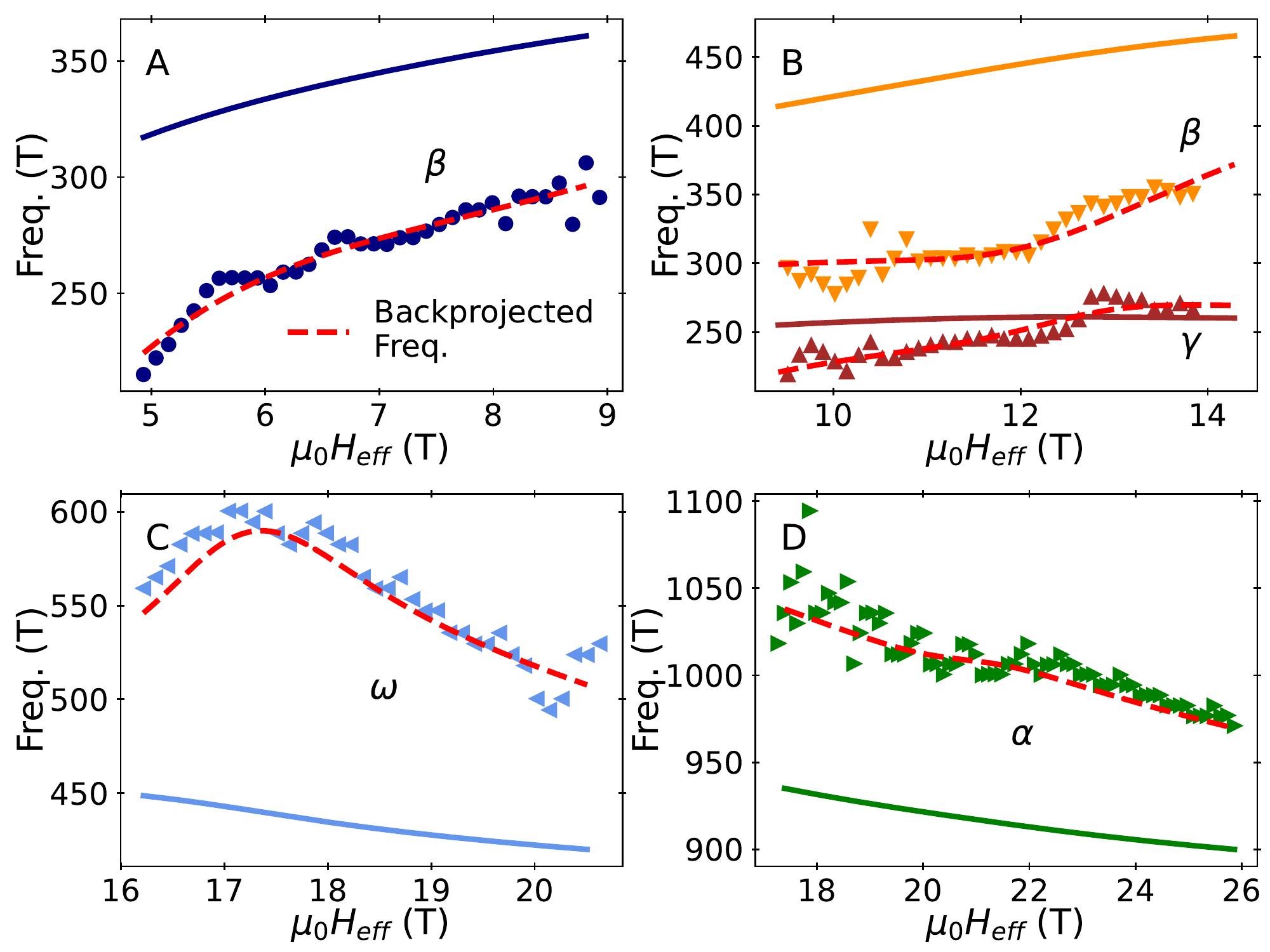}
	\caption{\label{fig:backprojection_down} Alternative possible true and backprojected frequencies: 
	For each region A to D, we show the measured data ($F_{\text{obs}}$) as scatter points. A possible $F_{\text{true}}$ is depicted as a solid line and its backprojection is shown as the dashed red line. }
\end{figure}

In the main paper, we discuss the difference between the oscillation frequency measured in a dHvA experiment $F_{\text{obs}}$, and the true 
frequency $F_{\text{true}}$. Fig. 4 shows possible true frequencies for regions $A-D$. 
The curves shown in Fig. 4 are not the only possibility for $F_{\text{true}}$, however. They are based on our assignment of measured dHvA frequencies to specific Fermi surfaces after comparison with the bandstructure and Fermi surface calculations. If our assignment is not correct, and the measured frequencies arise from different Fermi surfaces, the magnetic field dependence of the true frequencies may be correspondingly different. Fig. \ref{fig:backprojection_down} shows examples of alternative true frequencies that could also reproduce our $F_{\text{obs}}$.
We show these alternative possibilities for $F_{\text{true}}$ 
specifically to illustrate that if we have only the dHvA experiments, we have 
no information about whether the measured frequencies correspond to orbits 
on electron or hole pockets of the Fermi surface, or on majority- or 
minority-spin surfaces.

\section{WIEN2k calculations}
Using WIEN2k \cite{Blaha} we calculated the bandstructure of UCoGe with and without spin-polarization. For both calculations we used a $k$-mesh of 30000 $k$-points, the PBE-GGA exchange-correlation potential \cite{Perdew}, and we added spin-orbit coupling in the (001) direction. The energy separation between the core states and the valence states was taken to be -8.7 Ry, to ensure that the core charge was not leaking out. We took an RKmax of 9. We converged to an energy of 0.0001 Ry. As a comparison, in another U-based heavy fermion system, \chem{UPt_3}, it is known that the uncertainty of the Fermi energy with respect to the experimental determined value is $\sim 1$ mRy \cite{McCollam2021}. 

\subsection{Bandstructures and predicted dHvA frequencies}
\begin{figure*}
	\centering
	\begin{subfigure}[t]{0.348\textwidth}
		\includegraphics[width=\textwidth]{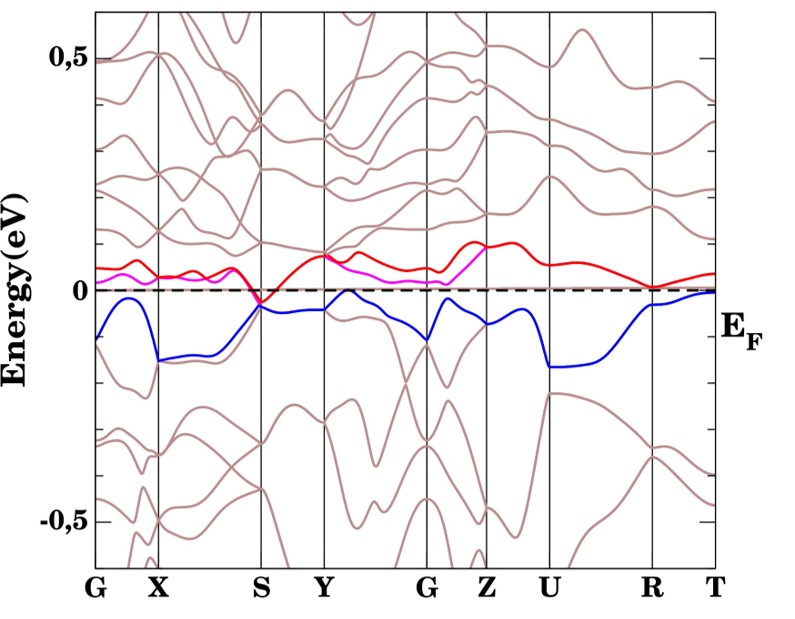}
		\caption{Bandstructure obtained from the paramagnetic calculation. The two-fold degenerate bands that cross the Fermi energy are color-coded: hole bands 243-244 (blue), electron bands 245-246 (magenta) and electron bands 247-248 (red).}
		\label{fig:BS_non-pol}
	\end{subfigure}
	~ 
	\begin{subfigure}[t]{0.35\textwidth}
		\includegraphics[width=\textwidth]{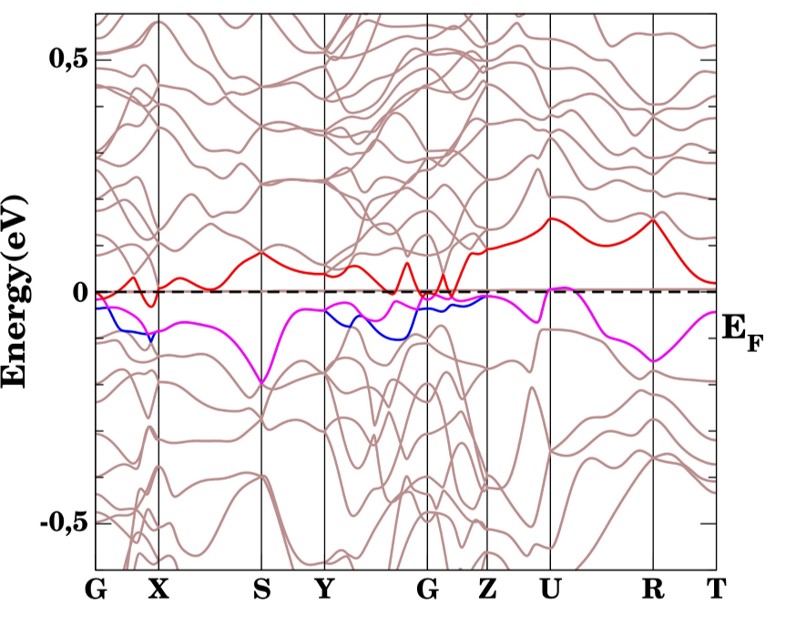}
		\caption{Bandstructure obtained from the spin-polarized calculation. Three bands cross the Fermi level: two hole bands 243 (blue) and 244 (magenta), and an electron band 245 (red).}
		\label{fig:BS_spin-pol}
	\end{subfigure}
	~ 
	\begin{subfigure}[t]{0.2\textwidth}
	\includegraphics[width=\textwidth]{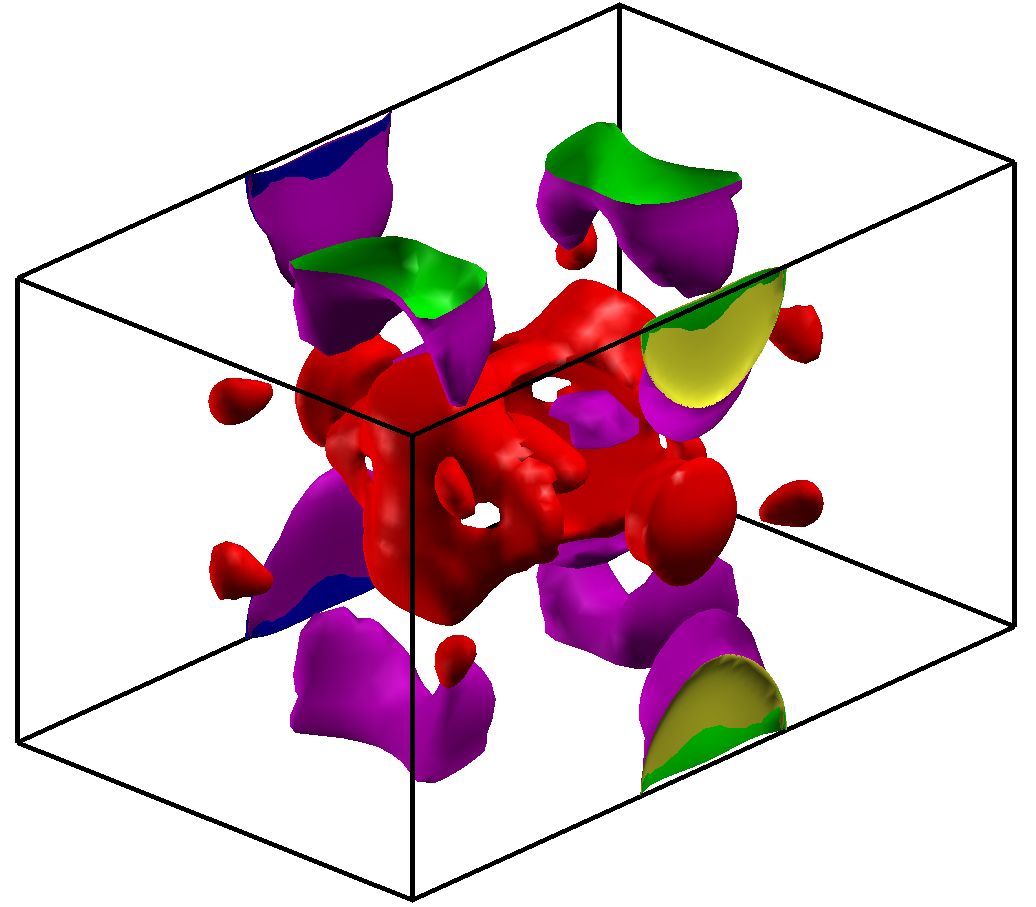}
	\caption{Fermi surface obtained from the spin-polarized calculation, showing band 243 (blue-yellow), 244 (magenta-green) and 245 (red-light-blue). }
	\label{fig:FS_spin-pol}
	\end{subfigure}
	\caption{The bandstructure of UCoGe for (a) the paramagnetic calculation and (b) the spin-polarized calculation, and (c) the Fermi surface for the spin polarized calculation. }\label{fig:bandstructure}
\end{figure*}

The bandstructures for the spin-polarized and non spin-polarized (paramagnetic) calculation can be seen in Fig. \ref{fig:bandstructure}.
Fig. \ref{fig:BS_non-pol}, shows the bandstructure from the paramagnetic calculation, where three two-fold degenerate bands cross the Fermi level: one hole band 243-244 (indicated in blue) and two electron bands, 245-246 (magenta) and 247-248 (red). Fig. \ref{fig:BS_spin-pol}, shows the bandstructure of the spin polarized calculation. Here, three bands cross the Fermi level: two hole bands, shown in blue (band 243) and magenta (band 244); and an electron band shown in red (band 245). The corresponding Fermi surface for the spin-polarized calculation is depicted in Fig. \ref{fig:FS_spin-pol}. The Brillouin zone is the same as indicated in Fig. \ref{fig:FS_PM_B=150_dn}.

The spin polarized Fermi surface is very different from the paramagnetic Fermi surface shown in the main paper (Fig. 3), but the similarity between the two is that they both have a complex multiband Fermi surface with small pockets. The bands around the Fermi energy are quite flat, and thus the Fermi surface is very sensitive to small shifts in $E_F$.

The paramagnetic Fermi surface presented in Fig. 3 of the main paper is similar to the non-magnetic Fermi surface presented by Samsel-Czeka\l a {\em et al.}\cite{Samsel-Czekala}, but there are some differences to the results of Fujimori {\em et al.}\cite{Fujimori}. We assume that the different results in reference \cite{Fujimori} are due to differences in the method, but very little information is given about the details of the calculation. 

The spin-polarized Fermi surface shown in Fig. \ref{fig:FS_spin-pol} is quite different to the ferromagnetic Fermi surface presented by Samsel-Czeka\l a {\em et al.}\cite{Samsel-Czekala} who used the LSDA exchange correlation potential instead of the GGA-PBE potential that we employed. 
Fig. \ref{fig:FS_spin-pol} is also very different to the Fermi surface presented by de la Mora and Navarro \cite{Mora2008}. Their calculations include the Hubbard-$U$, 
thereby localizing the uranium 5$f$ orbital. The spin-up and spin-down sheets of the Fermi surface were shown separately in reference \cite{Mora2008}, suggesting a non-fully-relativistic convention, such that the Fermi surface is naturally different from the one that we calculate\cite{Samsel-Czekala, Mora2008}.

\begin{figure}
	\includegraphics[width = 1.\linewidth]{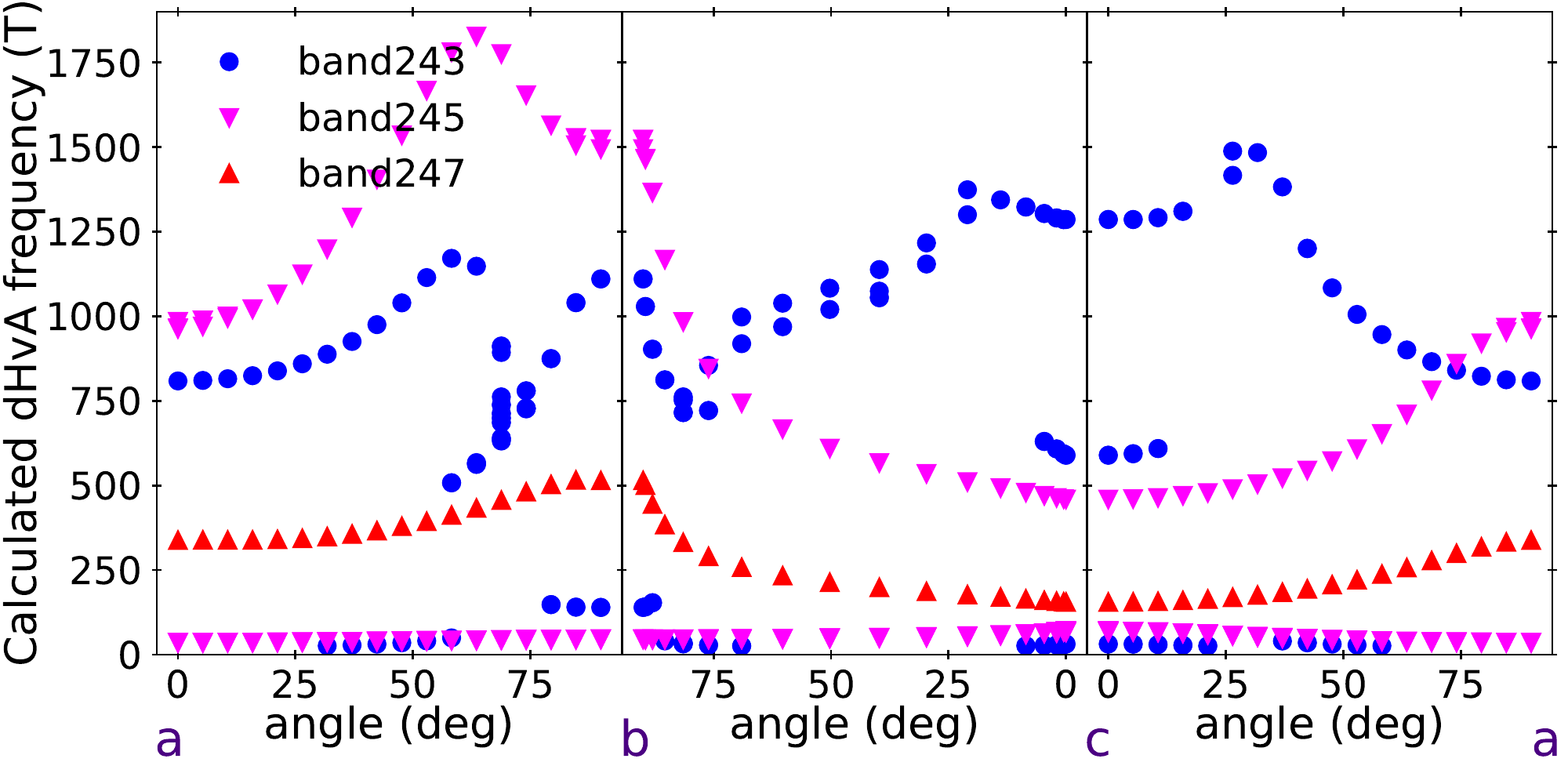}
	\caption{\label{fig:dHvA_calc} The expected dHvA frequencies as a function of magnetic field orientation for the paramagnetic calculation. It shows the dHvA frequencies of the three two-fold degenerate bands: band 243-244 (blue circles), 245-246 (magenta downward triangles) and 247-248 (red upward triangles). The principal axes are indicated at the bottom of the figure.  }
\end{figure}

\begin{figure}
	\includegraphics[width = 1.\linewidth]{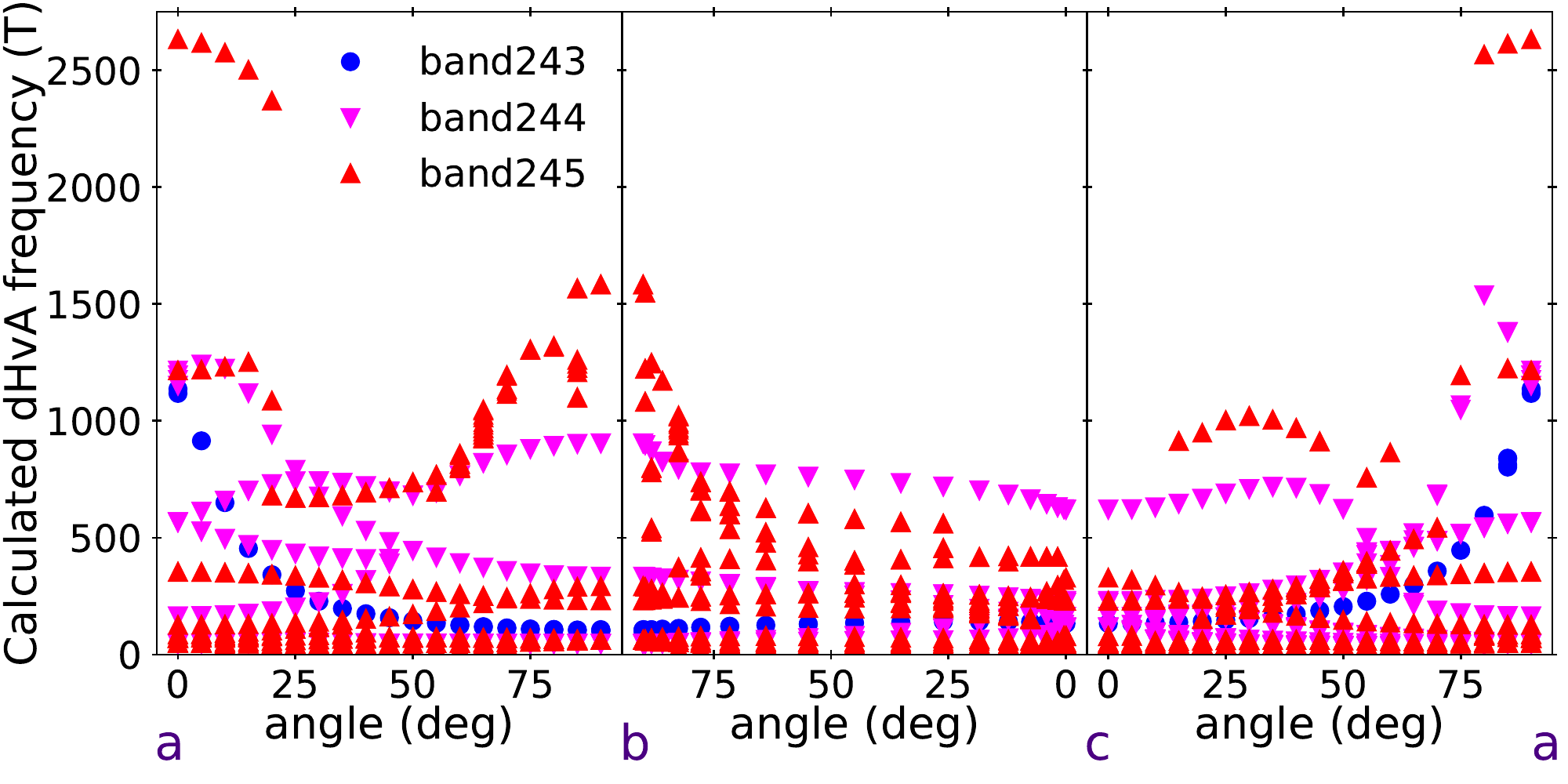}
	\caption{\label{fig:dHvA_calc_pol} The expected dHvA frequencies as a function of magnetic field orientation for the spin-polarized calculation. It shows the dHvA frequencies of the three bands that cross the Fermi level: band 243 (blue circles), 244 (magenta downward triangles) and 245 (red upward triangles). The principal axes are indicated at the bottom of the figure. }
\end{figure}

The angle dependence in magnetic field of the dHvA frequencies corresponding to the calculated Fermi surfaces \cite{Rourke} are shown in Fig. \ref{fig:dHvA_calc} (paramagnetic calculation) and Fig. \ref{fig:dHvA_calc_pol} (spin polarized calculation). 
The influence of magnetic field is not taken into account in the determination of these frequencies.

The dHvA frequencies are relatively low, and reflect the small Fermi surface pockets. 
The experimentally measured frequencies are shown in Fig. 2 of the main paper, and are approximately $F_\alpha = 1$~kT, $F_\omega = 580$~T, $F_\beta = 310$~T and $F_\gamma = 240$~T, with a weak magnetic field dependence in each case. The experiments were all conducted with magnetic field aligned along the crystallographic $c$-axis. 
Looking at Fig. \ref{fig:dHvA_calc} (see also table \ref{tab:effective_mass_calculated}) we can see that the predicted frequencies for $H\|c$ on the paramagnetic Fermi surface are $\sim 1.29$~kT, 590~T, 460~T, 156~T, and two much lower frequencies at $\sim 70$~T and 34~T.
In the spin-polarized case (Fig. \ref{fig:dHvA_calc_pol}), the predicted $c$-axis frequencies are $\sim 650$~T, and a cluster of many ($>10$) closely spaced frequencies below 300~T. 
The number of observed frequencies and their values, are therefore in better agreement with the predictions from the paramagnetic calculation.

Moreover, angle dependence of the measured dHvA frequencies was reported 
by Bastien {\em et al}. \cite{Bastien2016}, and is also in reasonable agreement with the predicted angle dependence for the paramagnetic Fermi surface, shown in Fig. \ref{fig:dHvA_calc}. The measured angle dependence of $F_\alpha$ \cite{Bastien2016} is slightly weaker than the highest frequency for $H||c$ shown in Fig. \ref{fig:dHvA_calc}, but has a similar trend.

We therefore tentatively match our measured dHvA frequencies 
with the quasiparticle orbits on the paramagnetic Fermi surface that have the closest predicted frequencies:
we match $F_\alpha$ with the large hole orbit on band 243-244 that has a predicted frequency of 1.29~kT; $F_\omega$ with the hole orbit, also on band 243-244, that has a predicted frequency of 590~T; $F_\beta$ with the electron orbit on band 245-246 that has a predicted frequency of 460~T; and $F_\gamma$ with the electron orbit on band 247-248 that has a predicted frequency of 156~T. 

Table \ref{tab:effective_mass_calculated} lists the calculated dHvA frequencies obtained from the paramagnetic Fermi surface for a magnetic field along the crystallographic $c$-axis. The table also shows the calculated band mass $m_{\text{band}}$ and curvature of the pockets. The values are obtained without adding the Zeeman-like term to spin-split the bands. For the predicted field-evolution of the spin-split dHvA frequencies, the reader is referred to Fig. 3 (c) in the main paper.

\begin{table}
	\begin{tabular}{l|l|l|l}
		band & Frequency (kT)&	$m_{\text{band}}$ ($m_e$) & curvature (kT\AA$^2$)\\
		\hline
		\hline
		243-244 & 0.034& 5.584 $\pm$	0.014 & -51.675$\pm$5.5836\\
		& 0.5905 & 7.571 $\pm$ 0.012 & 35.232 $\pm$ 1.124\\
		& 1.289& 13.914 $\pm$ 0.105 & -211.785 $\pm$ 3.063\\
		\hline
		245-246 & 0.071& 4.059 $\pm$ 0.072 & -179.77 $\pm$ 7.65\\
		& 0.4596 & 3.1629$\pm$	0.0146 & -14.733 $\pm$ 0.2732\\
		\hline
		247-248 & 0.1559& 1.2329$\pm$0.0011 & -8.1681 $\pm$ 0.04067\\
	\end{tabular}
	\caption{The calculated effective mass and curvature of the bands from the paramagnetic calculation for magnetic field along the crystallographic $c$-axis. All values are taken without including the model for applied external magnetic field.}\label{tab:effective_mass_calculated}
\end{table}

Comparison of the band masses given in table \ref{tab:effective_mass_calculated} with the effective masses $m^*$ extracted from our experimental data (Fig. \ref{fig:m_and_A_vs_B} a) shows relatively weak renormalisation of the quasiparticles on the higher frequency orbits (hole bands 243-244). For example, the measured $m^*$ for $F_\alpha$ varies between $\sim 11.5$ and 14~$m_e$ (see Fig. \ref{fig:m_and_A_vs_B} and the representative values in Table \ref{tab:effective_mass}). Within the experimental error bars, this suggests little or no mass enhancement over the predicted band mass of 13.9 $m_e$ for the 1.289~kT frequency shown in Table \ref{tab:effective_mass_calculated}.
The effective masses corresponding to $F_\beta$ and $F_\gamma$, which we assign to the lower frequency electron pockets (electron bands 245-246 and 247-248) are more strongly renormalised: a factor of $\sim 3$ for $F_\beta$, and a factor of $\sim 7$ for $F_\gamma$.

\subsection{Estimation of the $g$-factor}

As described in the main paper, our model to incorporate 
magnetic field dependence of the Fermi surface is based
on weak zero-field splitting of the bands and simple linear spin-splitting with increasing magnetic field. 
We use a Zeeman-like term $(\pm \frac{1}{2} g_0 \mu_B B_{\text{nom}})$, where
$B_{\text{nom}} = g/g_0 (B_{\text{ex}} + B_{\text{appl}})$ captures  
the ratio of the real $g$-factor to the free electron value, the intial exchange field $B_{\text{ex}}$, and the experimentally applied 
magnetic field $B_{\text{appl}}$.
We estimate the initial exchange field as  
$B_{\text{ex}} = 0.06 \mu_B \ \mu_0  N_A/V_M$, 
where 0.06 $\mu_B$ is the measured zero field moment \cite{Huy2008, Aoki2019}, $N_A$ is Avogadro's number and $V_M$ is the molar volume of UCoGe.
The resulting value of $B_{\text{ex}} = 0.0134$~T 
is sufficiently small that the dominant contribution to $B_{\text{nom}}$ is $g/g_0 B_{\text{appl}}$. 
This allows us to make an estimate of the $g$-factor in the case that we can associate some value of $B_{\text{nom}}$ with a specific applied magnetic field value.

\subsection{Lifshitz transitions at very high $B_{\text{nom}}$}
\begin{figure}[h]
	\centering
	\begin{subfigure}[t]{0.45\linewidth}
		\includegraphics[width=\linewidth,trim=5 5 5 5,clip]{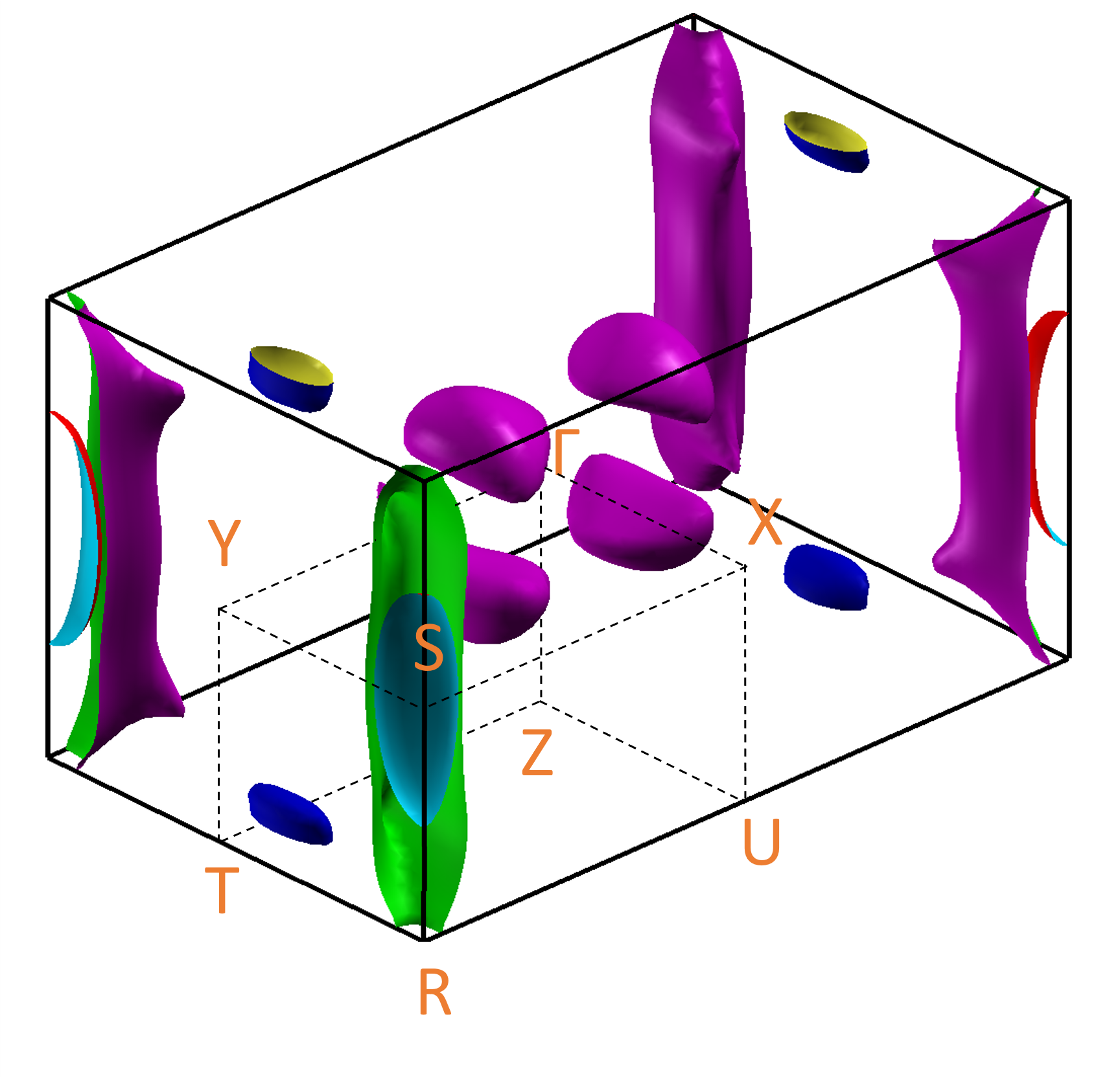}
		\caption{The calculated majority-spin Fermi surface of UCoGe for the $B_{\text{nom}}=150$~T. }
		\label{fig:FS_PM_B=150_dn}
	\end{subfigure}
	~ 
	\begin{subfigure}[t]{0.45\linewidth}
		\includegraphics[width=\linewidth,trim=5 5 5 5,clip]{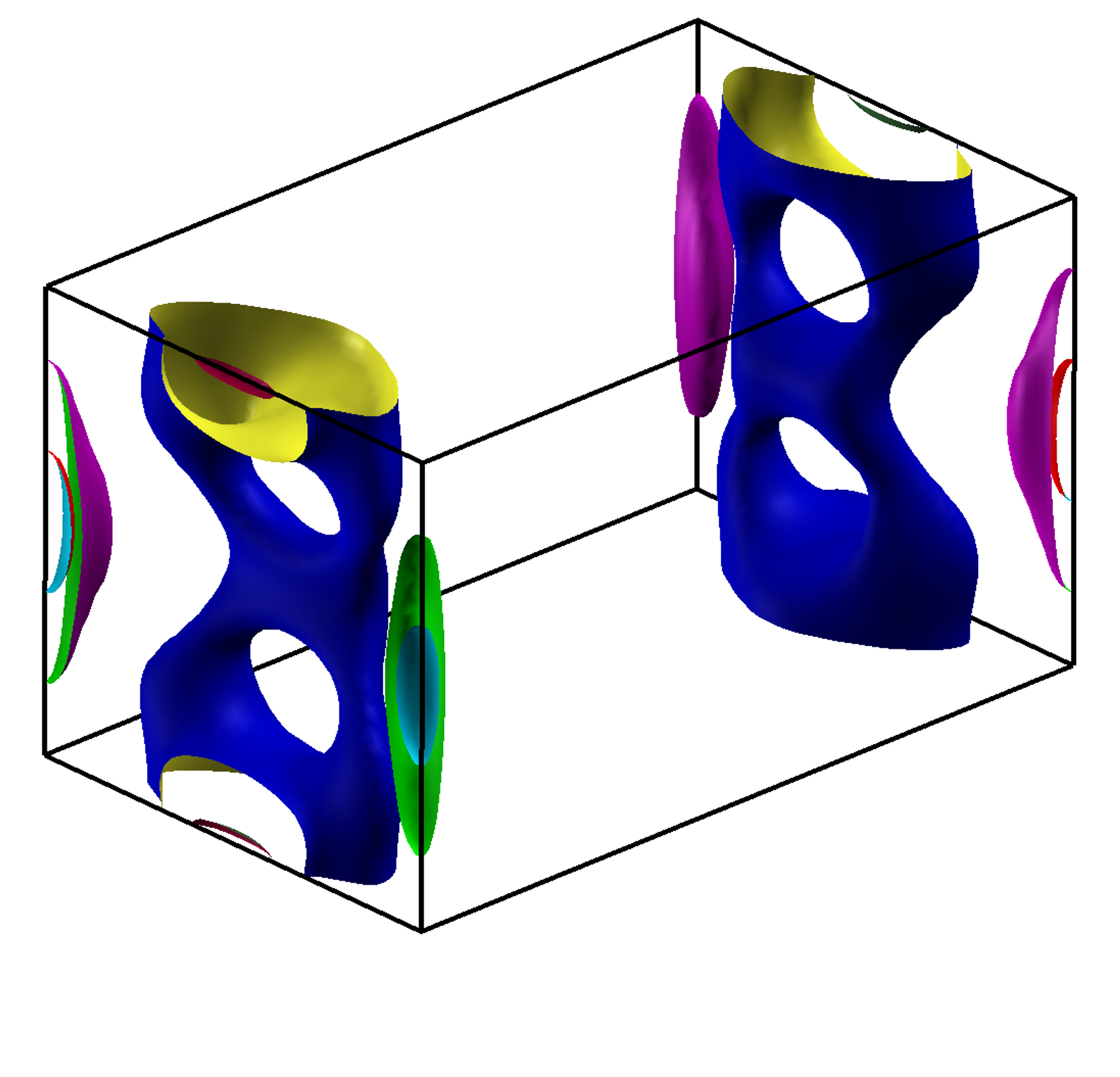}
		\caption{The calculated minority-spin Fermi surface of UCoGe for the $B_{\text{nom}}=150$~T.}
		\label{fig:FS_PM_B=150_up}
	\end{subfigure}
	\caption{The calculated paramagnetic Fermi surface of UCoGe with a splitting of $B_{\text{nom}} = 150$~T, \ref{fig:FS_PM_B=150_dn} shows the majority-spin Fermi surface and \ref{fig:FS_PM_B=150_up} shows the minority-spin Fermi surface. The hole-like band (number 243-244) is indicated in blue/yellow, the two electron-like bands are indicated in purple/green (bands 245-246) and red/light blue (bands 247-248). There are several changes in Fermi surface shape leading to new quasiparticle orbits, as one example, a new hole pocket appears from band 241-242 shown in dark red/dark blue around the T-point of the minority-spin Fermi surface.  }\label{fig:FS_PM_B=150}
\end{figure}

We discuss in the main paper that our calculations do not show significant Lifshitz transitions due to Fermi surface pockets that appear as a function of magnetic field  in the region of $B_{\text{nom}}<100$~T. However, at higher $B_{\text{nom}}$, the shape of the Fermi surface changes considerably. We show the Fermi surface for $B_{\text{nom}}=150$~T in Fig. \ref{fig:FS_PM_B=150}. Comparison with Fig. 3 in the main paper shows major changes of shape, including new orbits, occur near the R-S-Y-T region of the Brillouin zone.
\vspace{3cm}\\

